\documentclass[aps,prl,twocolumn,superscriptaddress,floatfix,nofootinbib,10pt]{revtex4-2}
\usepackage{graphicx}
\usepackage{amsmath,amssymb,tikz,physics,mathtools,bm}
\usepackage[T2A]{fontenc}
\usepackage[colorlinks, allcolors = blue, bookmarks=false]{hyperref}
\graphicspath{{img/}}

\newcommand{\lp}{\left (}
\newcommand{\rp}{\right )}

\newcommand{\lc}{\left[}
\newcommand{\rc}{\right]}

\begin{document}
\title{Charge-Informed Quantum Error Correction}

\author{Vlad Temkin}
\thanks{%
\href{mailto:vlad.temkin@berkeley.edu}{vlad.temkin@berkeley.edu}\\
\href{mailto:zackmw@caltech.edu}{zackmw@caltech.edu}\\
V.T. and Z.W. contributed equally to this work.
}
\affiliation{Department of Physics, University of California, Berkeley, California 94720, USA}

\author{Zack Weinstein}
\thanks{%
\href{mailto:vlad.temkin@berkeley.edu}{vlad.temkin@berkeley.edu}\\
\href{mailto:zackmw@caltech.edu}{zackmw@caltech.edu}\\
V.T. and Z.W. contributed equally to this work.
}
\affiliation{Department of Physics, University of California, Berkeley, California 94720, USA}
\affiliation{Department of Physics and Institute for Quantum Information and Matter, California Institute of Technology, Pasadena, California 91125, USA}

\author{Ruihua Fan}
\affiliation{Department of Physics, University of California, Berkeley, California 94720, USA}

\author{Daniel Podolsky}
\affiliation{Physics Department, Technion, Haifa 32000, Israel}

\author{Ehud Altman}
\affiliation{Department of Physics, University of California, Berkeley, California 94720, USA}
\affiliation{Materials Sciences Division, Lawrence Berkeley National Laboratory, Berkeley, California 94720, USA}

\date{\today}
 
\begin{abstract}
We investigate the statistical physics of quantum error correction in U(1) symmetry-enriched topological quantum memories. Starting from a phenomenological error model of charge-conserving noise, we study the optimal decoder assuming the local charges of each anyon can be measured. The error threshold of the optimal decoder corresponds to a continuous phase transition in a disordered two-dimensional integer loop model on the Nishimori line. Using an effective replica field theory analysis and Monte Carlo numerics, we show that the optimal decoding transition exhibits Berezinskii-Kosterlitz-Thouless universality with a modified universal jump in winding number variance. We further generalize the model beyond the Nishimori line, which defines a large class of suboptimal decoders. At low nonzero temperatures and strong disorder, we find numerical evidence of a disorder-dominated loop-glass phase which corresponds to a ``confidently incorrect'' decoder. The zero-temperature limit defines the \emph{minimum-cost flow} decoder, which serves as the ${\rm U}(1)$ analog of minimum-weight perfect matching in $\mathbb{Z}_2$ topological codes. Both the optimal and minimum-cost flow decoders are shown to dramatically outperform the charge-agnostic optimal decoder in symmetry-enriched topological codes. 
\end{abstract}

\maketitle

% Acronyms: FQH, BKT, SWSSB, RBIM, WA, NL, MWPM, MCF

Quantum error correction is a major focus of current research as an essential step for fault-tolerant quantum computation \cite{shor1995QEC, steane1996QEC, gottesman2010QEC, calderbank1996GoodQEC, terhal2015QECreview}. Topological codes such as the toric code \cite{kitaevTC2003, bravyi1998SurfaceCode, dennisTQM2002} are among the leading candidates for near-term implementation \cite{google2021TC, google2025dynamicsurfacecodes, google2023Suppressing, ibm2023d3SurfaceCode, eth2022d3SurfaceCode, bluvstein2025FaultTolerantArchitectures, bluvstein2024logical}, owing to their high error thresholds and simple spatial geometry.

Some of the naturally arising topologically ordered phases, such as the fractional quantum Hall (FQH) states \cite{tsui_two-dimensional_1982,laughlinState1983,willett_observation_1987,mooreReadState1991}, are also expected to encode quantum information robustly in the presence of decoherence \cite{das-sarma_topologically_2005,bravyi_universal_2006,freedman_towards_2006,nayakReview2008,alicea_new_2012,zijian2025FQHdec}. In these states, anyons carry an electrical charge in addition to their anyonic quantum numbers. Measuring this electrical charge provides significantly more information than measuring the anyon content alone. This additional information opens the door to new ``charge-informed'' decoders, which can potentially outperform ``charge-agnostic'' decoders not utilizing this information. 

\begin{figure}[h!]
     \centering
        \includegraphics[width=0.48\textwidth]{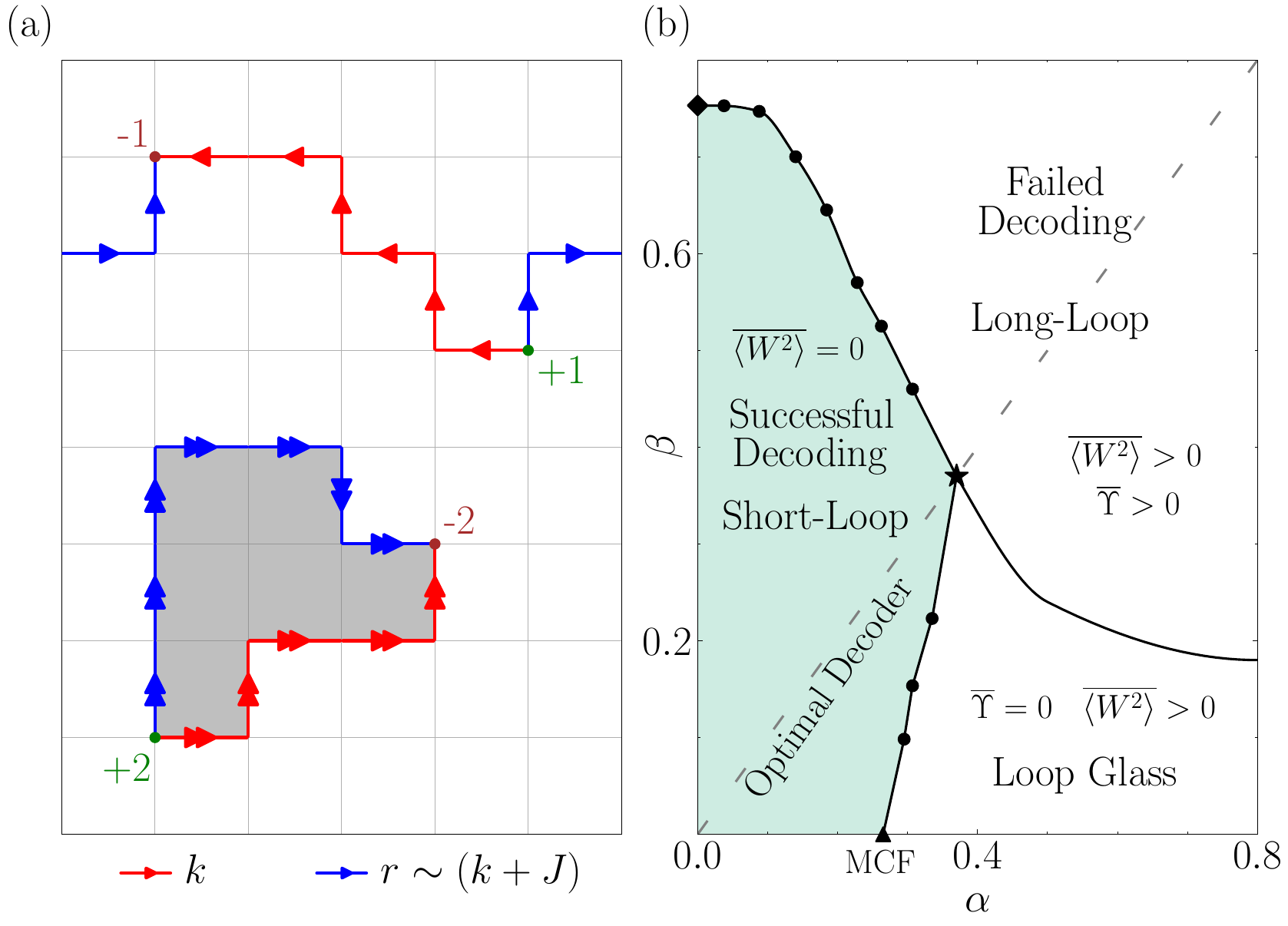}
        \caption{(a) Visualization of the quantum error correction problem investigated in this work. U(1) charges are created at the endpoints of an error configuration $k$ (red); an observer measures these charges, and annihilates them with a correction $r$ (blue). The combined error and correction form a collection of closed loops $J \sim (k - r)$. The bottom loop is topologically trivial and results in a successful recovery, while the top loop exhibits nontrivial winding and causes a logical error. (b) Numerical phase diagram of the model \eqref{eq:partition_fn} with disorder strength $\alpha$ and temperature $\beta$; the optimal decoder is described by the Nishimori line $\alpha = \beta$ (gray dashed line). Black dots denote numerically estimated phase boundaries, with error bars omitted. The phase boundary between the loop-glass and the long-loop phases is conjectured.}
        \label{fig:fig1}
\end{figure}

In this Letter, we develop the theory of quantum error correction in ${\rm U}(1)$ symmetry-enriched topological codes under charge-conserving noise. We assume a particular minimal model of decoherence which creates local quasiparticle-quasihole pairs; keeping the FQH states in mind, such a decoherence model roughly corresponds to low-frequency noise performed well below the Landau level gap, where anyonic quasiparticles are the natural local excitations of the ground state. 

Given this decoherence model, we study properties of the optimal decoder by a statistical physics mapping to a disordered integer loop model, with a Nishimori condition \cite{nishimori1981InternalEnergy, nishimoriBook2001} relating the disorder strength to the temperature. This model exhibits a short-loop phase and a long-loop phase, corresponding to phases of high and low decoding probabilities respectively, separated by a continuous phase transition. Using a combination of analytical field-theoretic arguments and detailed Monte Carlo numerics, we show that this decoding transition exhibits a modified Berezinskii-Kosterlitz-Thouless (BKT) universality \cite{berezinskii1971destruction,kosterlitzBKT1973,nelson1977Stiffness,jose_renormalization_1977,chaikin2013CMbook} with a universal jump in winding number variance of $1/\pi$, half the value of the clean loop model. Besides quantum error correction, we also investigate the possibility of U(1) strong-to-weak spontaneous symmetry breaking (SWSSB) --- a recently introduced mixed-state phenomenon describing the breakdown of a quantum (strong) symmetry to an ensemble (weak) symmetry \cite{leeJY2024SWSSB, makinde2023HydroSWSSB,lessaSWSSB2025,sala_spontaneous_2024}.

We further generalize the loop model by relaxing the Nishimori condition, allowing the temperature and disorder strength to be tuned independently, which can be interpreted as defining classes of suboptimal decoders \cite{zdeborova_statistical_2016,kim2025measurementinducedphasetransitionsquantum}. In the limit of weak disorder, field-theoretic arguments and corresponding numerics suggest that the BKT transition of the clean loop model is preserved, albeit with a disorder-dependent nonuniversal jump in the winding number variance. Conversely, the zero-temperature limit corresponds to a suboptimal \emph{minimum-cost flow} \cite{ahujaNetwork1993} decoder, which is a natural U(1) analog of minimum-weight perfect matching (MWPM) \cite{edmonds1965MWPM,edmonds1965MWPM1,dennisTQM2002,wangZ2Gauge2003} in $\mathbb{Z}_2$ topological codes. In contrast to the latter case, we find numerical evidence that a ``confidently incorrect'' decoder at large error strength extends to nonzero temperatures, corresponding to a disorder-dominated loop-glass phase in the loop model.

{\it Model.}---
As a phenomenological model of the charged anyons in our system, we consider an $L \times L$ periodic square lattice of quantum rotors. Each $i$th rotor has an integer-valued quasiparticle occupation number $\hat{n}_i$, and a phase operator $e^{i \hat{\varphi}_i}$ which creates and destroys quasiparticles via the commutation relations $[\hat{n}_i, e^{\pm i \hat{\varphi}_i}] = \pm e^{\pm i \hat{\varphi}_i}$. Our initial state is the quasiparticle vacuum $\rho_0 \equiv \dyad{0}^{\otimes N}$ 
with zero quasiparticles on each site, which is symmetric under the U(1) symmetry generator $\hat{\mathcal{U}}(\chi) \equiv \prod_i e^{-i \chi \hat{n}_i}$. Since the anyons exhibit nontrivial statistics, the vacuum $\rho_0$ is actually one of several topologically degenerate states \cite{wen_ground-state_1990,kitaevTC2003}, although this feature is not explicitly incorporated into our lattice model \cite{fna}. 

Our crucial physical assumption is that local, low-frequency noise can only create local quasiparticle-quasihole pairs on top of our encoded state. A minimal decoherence channel of this form, acting on each nearest-neighbor link $(i\mu)$ connecting site $i$ to site $i+e_{\mu}$ ($\mu \in \qty{x,y}$), is given by
%
%\footnote{We note that this model was previously introduced in Ref.~\cite{lessaSWSSB2025} in the context of strong-to-weak spontaneous symmetry breaking (SWSSB), a recently introduced mixed-state phenomenon describing the breakdown of a quantum (strong) symmetry to an ensemble (weak) symmetry \cite{leeJY2024SWSSB, makinde2023HydroSWSSB,lessaSWSSB2025,sala_spontaneous_2024}. In the SM, we argue that the optimal decoding transition in our model necessarily coincides with a (quasi-)SWSSB transition, and provide numerical evidence of (quasi-)SWSSB at large error strengths.}
\begin{equation}
\label{eq:channel}
    \begin{split}
        \mathcal{E}_{i\mu}(\rho) &\equiv \sum_{k_{i\mu} \in \mathbb{Z}} p_{k_{i\mu}} ~e^{-i k_{i\mu} \Delta_\mu \hat{\varphi}_i} ~\rho~  e^{i k_{i\mu} \Delta_\mu \hat{\varphi}_i},
    \end{split}
\end{equation}
where $\Delta_{\mu} \hat{\varphi}_i \equiv \hat{\varphi}_{i + e_\mu} - \hat{\varphi}_{i}$ is the lattice derivative of $\hat{\varphi}_i$ in the $\mu$ direction. Crucially, such a decoherence model conserves the total quasiparticle charge in the system. For simplicity, we shall assume a Gaussian probability distribution $p_{k_{i\mu}} \propto e^{-k_{i\mu}^2 / 2\alpha}$, so that the decohered state $\rho_{\alpha} \equiv [\prod_{i\mu} \mathcal{E}_{i\mu}](\rho_0)$ is controlled by the single parameter $\alpha$. We note that this model was previously introduced in Ref.~\cite{lessaSWSSB2025} in the context of SWSSB.

We can regard the full channel $\mathcal{E} \equiv \prod_{i\mu} \mathcal{E}_{i\mu}$ as applying the total error $\hat{E}_k \equiv \prod_{i\mu} e^{-i k_{i\mu} \Delta_{\mu} \hat{\varphi}_i}$ with probability $P_k \equiv \prod_{i\mu} p_{k_{i\mu}}$. Each error $\hat{E}_k$ is therefore labeled by an integer-valued lattice vector field $k \equiv \qty{k_{i\mu}}$, which can be visualized as a directed graph of one-dimensional field lines with integer labels (see Fig.~\ref{fig:fig1}). Each such error introduces $n_i \equiv \Delta_{\mu} k_{i\mu}$ quasiparticles to the site $i$, where $\Delta_{\mu} k_{i\mu} \equiv k_{ix} + k_{iy} - k_{i-e_x,x} - k_{i-e_y,y}$ is the lattice divergence of $k_{i\mu}$. In other words, quasiparticles are created (destroyed) at the tails (heads) of each field line of $k_{i\mu}$ [Fig.~\ref{fig:fig1}(a)]. 

Given the decohered state $\rho_{\alpha}$, an optimal quantum error correction protocol proceeds by first measuring the full set of charges $n \equiv \qty{n_i}$ on every site. Using the ``syndrome'' $n$, we attempt to recover the encoded state by annihilating positive and negative charges in pairs, which is implemented by a recovery operator $\hat{E}_r^{\dag}$ for another lattice vector field $r \equiv \qty{r_{i\mu}}$ with the same divergence $\Delta_{\mu} r_{i\mu} = n_i$. The recovery $r$ will successfully correct the error $k$ if the divergenceless vector field $k_{i\mu} - r_{i\mu}$, which can be visualized as a collection of integer-valued directed loops [Fig.~\ref{fig:fig1}(a)], exhibits no nontrivial windings around the cycles of the torus. In this case, we can express $k_{i\mu} - r_{i\mu} = \varepsilon_{\mu \nu} \Delta_{\nu} h_{\tilde{i}}$ as the curl of an integer-valued height field $h_{\tilde{i}}$ on the dual lattice sites $\tilde{i}$, where the vector field lines are realized as equal-elevation contour lines of the height field.

Conversely, if $k_{i\mu} - r_{i\mu}$ exhibits nontrivial windings around the cycles of the torus, then $\hat{E}^{\dag}_r \hat{E}_k$ transports anyons around the torus and can cause a logical error~\cite{fnb}. In this case, $k_{i\mu} - r_{i\mu} = J_{i\mu}$ is a divergenceless vector field which exhibits nonzero winding numbers $W_{\mu} \equiv \frac{1}{L} \sum_{i} J_{i\mu} \in \mathbb{Z}$.

{\it Optimal Decoding.}--- The set of errors $k$ corresponding to a given syndrome $n$ can be organized into \textit{winding sectors}; while the winding of a single error is undefined, two errors $k$ and $k'$ with $\Delta_{\mu} k_{i\mu} = \Delta_{\mu} k'_{i\mu}$ belong to the same winding sector if $k_{i\mu} - k'_{i\mu}$ has zero windings around the torus. Any two errors in the same winding sector can be corrected by the same recovery operation. Therefore, the optimal ``maximum likelihood'' decoding protocol is to compute the total probability of each winding sector and apply a recovery which corrects the most likely sector~\cite{dennisTQM2002}.

We can obtain significant theoretical insight on the success rate of our optimal decoder via a statistical physics mapping to a disordered loop model. Given an error $k$, the total probability $Z_k$ of observing the charge distribution $n = \Delta k \equiv \qty{\Delta_{\mu} k_{i\mu}}$ is given by summing over all errors with the same divergence: 
\begin{equation}
\label{eq:partition_fn}
    \begin{split}
        Z_k &\equiv \sum_{J: \Delta J = 0} P_{k + J} \\
        &\propto \sum_{J : \Delta J = 0} \exp \qty{ - \frac{1}{2\alpha} \sum_{i\mu} (k_{i\mu} + J_{i\mu})^2 } ,
    \end{split}
\end{equation}
where the sum is performed over all integer-valued divergenceless vector fields $J$ (i.e., $\Delta_{\mu} J_{i\mu} = 0$). Note that $Z_k = Z_{k+J}$ is independent of the representative error $k$ for a given charge distribution $n$. We can similarly compute the total probability of the winding sector for $k$ by demanding that $J_{i\mu} = \varepsilon_{\mu \nu} \Delta_{\nu} h_{\tilde{i}}$ has zero winding.

The quantity $Z_k$ can be interpreted as a partition function of a disordered statistical physics model of directed closed loops $J$. In the Supplemental Material (SM) \cite{SOM}, we employ standard techniques to recast $Z_k$ as the partition function of a Villain XY model with imaginary disorder: 
\begin{equation}
\label{eq:partition_fn_XY}
    Z_k \propto \sum_m \int_{\theta} \exp \qty{ - \frac{\alpha}{2} \sum_{i\mu} (\Delta_{\mu} \theta_i - 2\pi m_{i\mu})^2 + i \sum_i n_i \theta_i } ,
\end{equation}
where $\theta_i \in [0, 2\pi)$ denotes an XY spin, $\int_{\theta} \equiv \int \prod_i \frac{\dd{\theta_i}}{2\pi}$, and $m \equiv \qty{m_{i\mu}}$ is an integer-valued gauge field on the links of the lattice. The disorder $\prod_{i} e^{i n_{i} \theta_{i}}$ can be regarded as a collection of frozen ``clock charges'' at the syndrome locations \cite{jose_renormalization_1977}. Upon restricting to a particular winding sector, $Z_k$ can alternatively be regarded as a disordered height model \cite{SOM}. Since the disorder $k$ is drawn from the same probability distribution $P_k$ which comprises the Boltzmann weights, these disordered models feature a Nishimori condition with enhanced replica symmetry akin to the Nishimori random-bond Ising model (RBIM) \cite{nishimoriBook2001,gruzberg2001RBIM}.

Our optimal decoding protocol will succeed with high probability if, for each typical error $k$, the relative probability of all other errors with the same charge distribution in different winding sectors is negligibly small. In other words, error correction is possible when the winding number $W_{\mu}$ is zero and nonfluctuating in the thermodynamic limit. This can be probed by the disorder-averaged mean-square winding number, $\overline{\expval{W_{\mu}^2}} \equiv \sum_k P_k \expval{W^2_{\mu}}_k$, where $\expval{\cdot}_k$ denotes a statistical average with respect to the partition function \eqref{eq:partition_fn} and the overline denotes the disorder average. Simple perturbative arguments suggest that $\overline{\expval{W_{\mu}^2}} \sim e^{-L/\xi}$ for small $\alpha$, while $\overline{\expval{W_{\mu}^2}} \sim \mathcal{O}(1)$ at large $\alpha$ indicates the failure of optimal decoding. 

In the Villain XY representation of $Z_k$, it is especially natural to examine the disorder-averaged \emph{helicity modulus} $\overline{\Upsilon} \equiv \sum_k P_k \Upsilon_k$, given in each disorder realization by the variance in winding number:
\begin{equation}
    \label{eq:helMod}
    \Upsilon_k \equiv \frac{1}{2} \sum_{\mu = x, y} \qty[ \expval{W_{\mu}^2}_k - \expval{W_{\mu}}^2_k ] .
\end{equation}
This quantity physically probes the phase rigidity of the XY spins \cite{pollock_path-integral_1987,xu_high-precision_2019,SOM}. In the clean model with $k_{i\mu} = 0$, $\Upsilon_0$ can be used to distinguish two phases: a disordered phase at small $\alpha$ with $\Upsilon_0 = 0$, and a ``superfluid'' phase at large $\alpha$ with $\Upsilon_0 \neq 0$. The two regimes are separated by a universal jump in $\Upsilon_0$ of $2/\pi$ at a BKT transition \cite{nelson1977Stiffness,chaikin2013CMbook}. It is natural to expect that a qualitatively similar transition arises in the disorder-averaged helicity modulus $\overline{\Upsilon}$ at a critical parameter $\alpha_c$. Furthermore, as shown in the SM \cite{SOM}, the Nishimori condition implies $\overline{\expval{W_{\mu}^2}} = 2 \overline{\expval{W_{\mu}}^2}$, and therefore the onset of nonzero $\overline{\Upsilon}$ at the transition also signals the failure of the optimal decoder.

Another important characterization of the BKT transition in the clean model is the behavior of two-point correlation functions $\expval{e^{i(\theta_x - \theta_y)}}_0$ between two sites $x$ and $y$, which transition from exponentially decaying in $\abs{x-y}$ at small $\alpha$ to algebraically decaying at large $\alpha$ \cite{kardarFields2007,chaikin2013CMbook}. As we show in the SM \cite{SOM}, the correlation function $\expval{e^{i(\theta_x - \theta_y)}}_k$ in the disordered case can be directly related to the R\'enyi-1 and fidelity correlators \cite{lessaSWSSB2025,weinsteinR12025,liu_diagnosing_2025a}, two recently proposed diagnostics of SWSSB. Specifically, the R\'enyi-1 correlator is given by
\begin{equation}
    \begin{split}
        R_1(x,y) &\equiv \tr[ e^{i (\hat{\varphi}_x - \hat{\varphi}_y)} \sqrt{\rho_{\alpha}} e^{-i (\hat{\varphi}_x - \hat{\varphi}_y)} \sqrt{\rho_{\alpha}} ] \\
        &= \sum_k P_k \sqrt{\expval{e^{i(\theta_x - \theta_y)}}_k} .
    \end{split} \label{eq:R1corr}
\end{equation}
Therefore, (quasi-)long-range order in $\expval{e^{i(\theta_x - \theta_y)}}_k$ for typical disorder realizations $k$ indicates that the strongly U(1)-symmetric mixed state $\rho_{\alpha}$ exhibits U(1) (quasi\mbox{-})SWSSB \cite{lessaSWSSB2025}. Since the algebraic decay of two-point correlations is precisely determined by the helicity modulus in the clean model \cite{chaikin2013CMbook}, it is natural to expect that any (quasi-)SWSSB transition must necessarily coincide with the optimal decoding threshold.

\begin{figure*}[t]
     \centering
         \includegraphics[width=\textwidth]{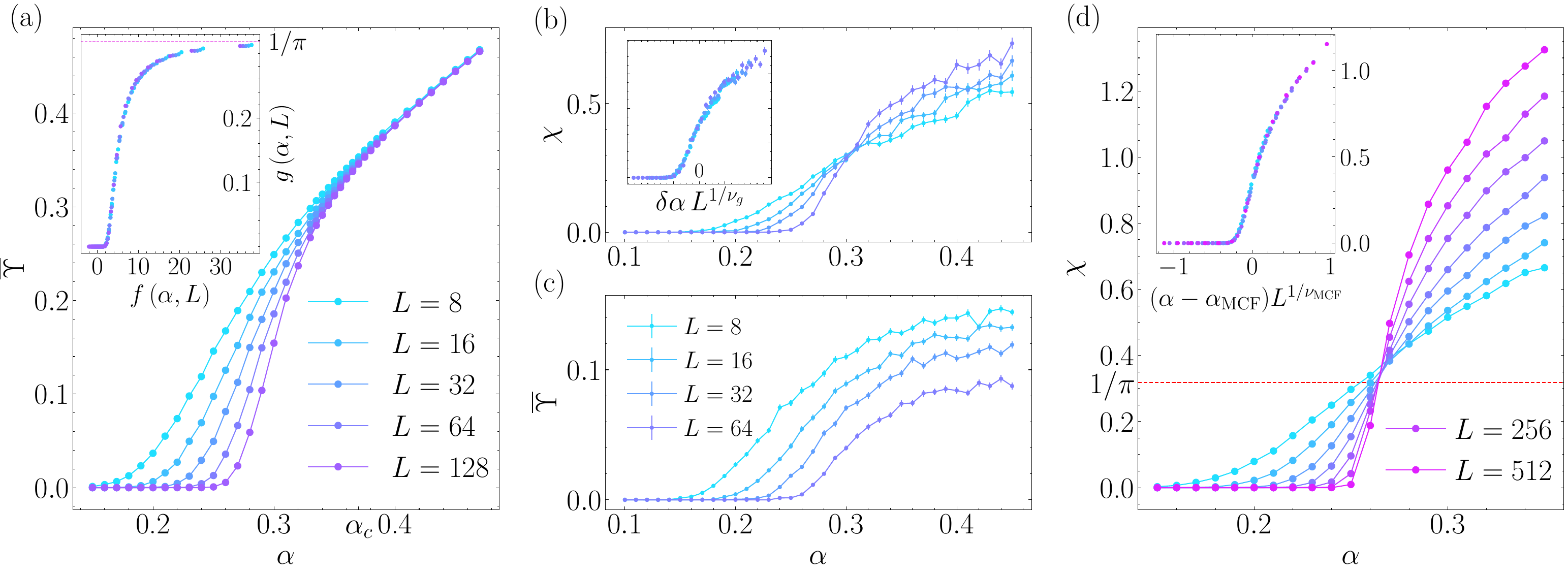}
        \caption{(a) Disorder-averaged helicity modulus $\overline{\Upsilon}$ on the Nishimori line as a function of the error strength $\alpha$, for various system sizes. Vanishing $\overline{\Upsilon}$ at small $\alpha$ indicates successful decoding, while nonvanishing $\overline{\Upsilon}$ at larger $\alpha$ implies a finite logical error probability. Inset: BKT finite-size scaling collapse with $f\lp \alpha, L\rp \equiv -\ln L + a \lp \alpha^{-1} - \alpha_c^{-1}\rp^{-1/2}$ and $g\lp \alpha, L\rp \equiv \overline{\Upsilon} \lc 1 +  \lp \ln L + c\rp^{-1}\rc^{-1}$ with $a = 7.6$, $c = 7.11$. This scaling collapse is consistent with a universal jump of $1/\pi$. (b) Edwards-Anderson helicity modulus $ \chi \equiv \frac{1}{2} \sum_{\mu} \overline{\langle W_{\mu} \rangle^2}$ on the line $\beta = 0.15$ as a function of the error strength $\alpha$, for various system sizes. Inset: finite-size scaling collapse with $\delta\alpha = \alpha - 0.305$ and $\nu_g = 2.5$. (c) Disorder-averaged helicity modulus $\overline{\Upsilon}$ on the $\beta = 0.15$ line. The clear crossing and scaling collapse in $\chi$ at $\alpha_g \approx 0.305$, combined with the systematic decrease in $\overline{\Upsilon}$ with increasing system size, is suggestive of a loop-glass phase for $\alpha > \alpha_g$. (d) Edwards-Anderson helicity modulus $\chi$ from the (linear) MCF decoder, corresponding to the $\beta = 0$ line. Inset: finite-size scaling collapse with $\alpha_{\text{MCF}} = 0.265$ and $\nu_{\text{MCF}} = 2.19$.}
        \label{fig:NL}
\end{figure*}

\emph{Field theory.}--- It is interesting to relax the Nishimori condition in the model~\eqref{eq:partition_fn_XY} by replacing $\alpha$ in the Boltzmann weights with a new independent parameter $\beta$. The optimal decoder is described by the Nishimori line (NL) $\alpha = \beta$, while other values of $\beta$ can be interpreted as defining suboptimal decoding protocols \cite{zdeborova_statistical_2016,SOM}. In the limit of large $\beta$ and weak disorder $\alpha$, we can formally coarse-grain the model \cite{SOM} to obtain a continuum field theory described by the action
\begin{equation}
\label{eq:field_theory}
    \mathcal{S}_k[\theta] \equiv \int \dd[2]{x} \qty{ \frac{K}{2} (\partial_{\mu} \theta)^2 + i k^{\mu} \partial_{\mu} \theta } ,
\end{equation}
where $\theta(x) \sim \theta(x) + 2\pi$ is a compact scalar field, $K$ is the helicity modulus, and $k^{\mu}(x)$ is a Gaussian random field with zero mean and variance $\overline{k^{\mu}(x) k^{\nu}(y)} = D \delta^{\mu \nu} \delta(x-y)$. For $\beta \gg 1$ and $\alpha \ll 1$, we can identify $K \simeq \beta$ and $D \simeq \sum_{k_{i\mu} \in \mathbb{Z}} p_{k_{i\mu}} k_{i\mu}^2 \approx e^{-1/2\alpha}$. By averaging over quenched disorder using the replica trick, several quantitative insights can be derived \cite{SOM}. For example, the replica field theory predicts that $\overline{\Upsilon}$ undergoes a nonuniversal jump of $\frac{2}{\pi} - D$ at the transition between the disordered and superfluid phases, while $\overline{\expval{W^2}} \equiv \frac{1}{2} \sum_{\mu} \overline{\expval{W_{\mu}^2}}$ undergoes a universal jump of $2/\pi$ as in the clean XY model.

While this field theory is formally derived in the weak-disorder limit, it is natural to postulate that it continues to describe the entire long-loop phase as an effective field theory, with renormalized parameters $K$ and $D$. If we assume that such a field theory describes the $\alpha > \alpha_c$ phase of the NL, then we immediately predict that $\overline{\Upsilon} = \frac{1}{2} \overline{\expval{W^2}}$ now exhibits a universal jump of $1/\pi$ at the critical point. This is consistent with the results of our numerics, as we shall now discuss.

{\it Numerical results.}--- We study the optimal decoder's error threshold by numerically computing the disorder-averaged helicity modulus \eqref{eq:helMod} using Monte Carlo simulations. First, we independently sample an error $k_{i\mu}$ with probability $p_{k_{i\mu}} \propto e^{-k_{i\mu}^2/2\alpha}$ on each link, and then we use the worm algorithm (WA) \cite{prokofievWorm2001} to stochastically sample closed loops $J_{i\mu}$ with relative probabilities given by the Boltzmann weights in Eq.~\eqref{eq:partition_fn}. During the simulation, WA naturally samples configurations from all winding sectors, and $\Upsilon_k$ is computed from the statistical variance of the winding number. The decoder succeeds if the zero-winding sector is sampled with the highest frequency, and fails otherwise. See the SM \cite{SOM} for additional details on WA.

As expected, we find a continuous phase transition in the behavior of the optimal decoder at a critical error strength $\alpha_c$ [Fig.~\ref{fig:NL}(a)]. The decoder succeeds with exceedingly high probability for $\alpha < \alpha_c$, where typical loops $J$ are short and the zero-winding sector is overwhelmingly dominant; this is the disordered phase of the XY spins. In contrast, the decoder fails with finite probability in the phase $\alpha > \alpha_c$, which exhibits large loops and a nonzero average helicity modulus.

To characterize the phase transition, we employ a BKT finite-size scaling ansatz and implement the procedure of Ref.~\cite{weberMCXY1988} to determine the critical point $\alpha_c$. At the BKT transition in a clean XY model, the helicity modulus obeys the following finite-size scaling ansatz \cite{SOM}: 
\begin{equation}
\label{eq:stiffness_FSS}
    \Upsilon_0\qty( \alpha_c,  L) = \Upsilon_0(\alpha_c,\infty) \qty(1 + \frac{\qty[\pi \Upsilon_0(\alpha_c, \infty)]^{-1}}{\log L + c}),
\end{equation}
where $\Upsilon_0(\alpha_c, \infty) = 2/\pi$ is the universal stiffness jump in the thermodynamic limit, and $c$ is a nonuniversal constant. From the effective field theory, we expect a similar scaling ansatz for $\overline{\Upsilon}(\alpha_c, L)$ with a modified stiffness jump $\overline{\Upsilon}(\alpha_c, \infty) = 1/\pi$ in our model. As in Ref.~\cite{weberMCXY1988}, we determine $\alpha_c$ by numerically fitting $\overline{\Upsilon}(\alpha, L)$ to the ansatz \eqref{eq:stiffness_FSS} for each $\alpha$, which minimizes the least-squares error with respect to the parameter $c$. $\alpha_c$ is then chosen as the value of $\alpha$ with the smallest least-squares error. Using this procedure, we find $\alpha_c \approx 0.37$ \cite{SOM}. We can then extend our finite-size scaling ansatz to $\alpha$ away from the critical point \cite{sandvik_computational_2010}, where we find a good scaling collapse [Fig.~\ref{fig:NL}(a), inset]. This behavior is consistent with a BKT transition with the reduced stiffness jump of $1/\pi$.

% \what{Change this paragraph}
WA also offers a natural method to compute the 2-point correlation function \cite{prokofievWorm2001}, which can be leveraged to determine the R\'enyi-1 correlator \eqref{eq:R1corr}. Numerical results are presented in the SM \cite{SOM}. In the short-loop phase, we observe $R_1$'s expected exponential decay. The behavior in the long-loop phase is substantially longer ranged, but strong disorder effects and finite lattice sizes prevent us from determining the precise decay of the correlator. Our numerical results are therefore suggestive, albeit not conclusive, of (quasi-)SWSSB for $\alpha > \alpha_c$.

Besides the NL $\alpha = \beta$, we also study transitions along several lines of the form $\beta = \lambda \alpha$ for various $\lambda$. The resulting phase diagram is shown in Fig.~\ref{fig:fig1}(b). In the weak disorder limit (i.e., $\lambda > 1$), we again find that the phase transition between short-loop and long-loop phases exhibits excellent BKT-like scaling. Using a two-parameter fit for both the stiffness jump $\overline{\Upsilon}(\alpha_c, \infty)$ and $c$ as in Ref.~\cite{weberMCXY1988}, we find that $\overline{\Upsilon}$ exhibits a nonuniversal jump at the transition which is reduced from $2/\pi$ with increasing $\alpha$. Additionally, we find that $\overline{\expval{W^2}}$ is consistent with a universal jump of $2/\pi$ at the transition~\cite{SOM}. Both of these observations are consistent with the predictions of the field theory~\eqref{eq:field_theory}.

Remarkably, in addition to the short-loop and long-loop phases found on and above the NL, we find numerical evidence which suggests an additional \emph{loop-glass} phase for large $\alpha$ below the NL [Fig.~\ref{fig:fig1}(b)]. This phase is characterized by a vanishing $\overline{\Upsilon}$ in the thermodynamic limit, but a nonvanishing Edwards-Anderson helicity modulus $\chi \equiv \frac{1}{2} \sum_{\mu} \overline{\expval{W_{\mu}}^{2}}$. Specifically, Figures~\ref{fig:NL}(b) and \ref{fig:NL}(c) depict $\overline{\Upsilon}$ and $\chi$ along the cut $\beta = 0.15$. $\chi$ exhibits a clear crossing at $\alpha_g \approx 0.305$, and a conventional (i.e., non-BKT) finite-size scaling collapse with correlation length exponent $\nu_g \approx 2.5$. In contrast, $\overline{\Upsilon}$ exhibits no such crossing and systematically decreases towards zero with increasing system size.

By interpreting models with $\beta \neq \alpha$ as suboptimal decoders, the loop-glass phase corresponds to a regime where the decoder is ``confidently incorrect'' \cite{SOM,kim2025measurementinducedphasetransitionsquantum}: although the winding number is nonfluctuating in individual disorder realizations, there is a nonvanishing probability that the resulting winding sector differs from that of the sampled error. While a similar phase exists at strictly zero temperature in the two-dimensional RBIM, our numerics suggest that this phase extends to nonzero $\beta$ in the present model.

{\it Minimum Cost Flow.}--- In $\mathbb{Z}_{2}$ topological codes, MWPM serves as an efficient suboptimal decoder when an efficient optimal decoder cannot be constructed \cite{dennisTQM2002}. In the present setting, access to the integer-valued charges of the errors allows for a natural generalization of MWPM using \textit{minimum-cost flow} (MCF) algorithms \cite{ahujaNetwork1993}. Similar to MWPM, MCF decoding constructs a recovery operation $\hat{E}^\dagger_r$ by choosing $r$ as the single most probable error consistent with the observed syndrome $n$. That is, $r$ is chosen to minimize the energy function $\frac{1}{2}\sum_{i\mu} r_{i\mu}^2 = \frac{1}{2}\sum_{i\mu} \qty(J_{i\mu} + k_{i\mu})^2$, which is a quadratic instance of the MCF problem \cite{ahujaNetwork1993}. This decoder can be interpreted as the $\beta = 0$ limit of the disordered loop model.

Although there exists a strongly polynomial algorithm for solving the quadratic MCF problem \cite{laszloQMCF2016}, it is difficult to implement and simulate numerically. Luckily, for all $\alpha$ near and below the optimal decoding threshold $\alpha_{c}$, the probability of sampling errors with $\abs{k_{i\mu}} > 1$ is quite small. Therefore, we can instead construct a suboptimal decoder which chooses $r$ to minimize $\sum_{i\mu} \abs{r_{i \mu}}$. This is a \textit{linear} instance of the MCF problem; it can be solved with standard integer linear programming methods \cite{ahujaNetwork1993}, and we expect it to behave essentially identically to the quadratic MCF problem in the present context.

To numerically study the error threshold of the MCF decoder, we sample independent errors $k_{i\mu}$ with probability $p_{k_{i\mu}}$ on each link, and then we find the linear MCF solution using an implementation by Google OR-Tools \cite{ortools} of the algorithm developed in Ref.~\cite{ahujaMCF1992}. We then compute the winding number $W_{\mu}$ of the MCF solution relative to the sampled error. By averaging over many sampled errors, we obtain the Edwards-Anderson helicity modulus $\chi$. Note that $\chi = \overline{\expval{W^2}}$ along the $\beta = 0$ line.

In contrast to the optimal decoder, $\chi$ within MCF is found to obey a more conventional finite-size scaling. We observe a crossing at the critical threshold $\alpha_{\text{MCF}} \approx 0.265$, and an excellent finite-size scaling collapse with a finite value of the correlation length exponent $\nu_{\text{MCF}} \approx 2.19$ [Fig.~\ref{fig:NL}(d), inset]. Interestingly, we find that the value of $\chi$ at the MCF critical point is consistent with the value $1/\pi$ [Fig.~\ref{fig:NL}(d), red dashed line], the value predicted analytically and observed numerically at the Nishimori critical point.

{\it Discussion.}--- We have investigated the optimal and suboptimal decoding of a topological code enriched with U(1) symmetry. For the phenomenological decoherence model considered here, the error threshold of the optimal decoder coincides with a continuous phase transition in a disordered integer loop model [Eq.~\eqref{eq:partition_fn}]. The general symmetry features of this model are insensitive to the specific choice of decoherence model, as long as charge is conserved. Consequently, we expect that a BKT-like decoding transition with a modified stiffness jump will describe a broad class of local and U(1)-symmetric decoherence models.

It is interesting to quantify the amount of additional information gained by measuring the U(1) charges of the anyons, beyond simply measuring their topological charges alone. In the SM \cite{SOM}, we consider the $\mathbb{Z}_{2}$ toric code enriched with a U(1) symmetry, such that the $e$ anyons carry global U(1) charge. In this setting, the ``charge-agnostic'' optimal decoder is described by the standard Nishimori RBIM \cite{dennisTQM2002} and the corresponding error threshold lies at $\alpha_{c}^{\text{RBIM}} \simeq 0.179$ \cite{SOM}. Therefore, both charge-informed optimal and MCF decoders dramatically outperform the charge-agnostic optimal decoder.

The decoders in the present work assume perfect measurements of each charge, and that the anyon's topological charge can be determined perfectly from the measured U(1) charge. This naturally raises the question of fault-tolerant decoding schemes in the presence of imperfect charge measurements. Additionally, we have assumed throughout that the U(1) charge is perfectly conserved. It would be interesting to investigate whether charge-informed decoders still strongly outperform charge-agnostic decoders when charge conservation is weakly broken.

Beyond its relevance to error correction, the statistical physics model introduced in this work exhibits a rich phase diagram featuring a broad ``critical'' phase and a putative finite-temperature loop-glass phase. While our numerical results on the NL and in the weak disorder limit $\alpha < \beta$ are consistent with a field-theoretic prediction of BKT-like transitions, much less is understood about the putative loop-glass phase and its nearby critical phenomena. Given the universal stiffness jump of $1/\pi$ on the NL, and the apparent winding number variance of $1/\pi$ at the critical point of the MCF decoder, it is tantalizing to guess that $\chi = 1/\pi$ everywhere along the transition between short-loop and loop-glass phases. Larger scale numerical studies are necessary to precisely determine the critical phenomena of transitions between short-loop and loop-glass phases, as well as between the loop-glass and long-loop phases.

{\it Note added.}--- While preparing this manuscript, we became aware of an independent related work \cite{fnc}.

% \begin{acknowledgments}
    {\it Acknowledgments.}--- We thank Zijian Wang for collaboration on a related project and fruitful discussions. We are grateful to Michael Levin, Zac Tobias, Nathaniel Selub, Jong Yeon Lee, and Akash Vijay for helpful discussions. This work was supported in part by the NSF QLCI program through Grant No. OMA-2016245, by a Simons Investigator Award (E.A.), by the Israel Science Foundation grant no. 2005/23 (D.P.), and by the Gordon and Betty Moore Foundation Grant GBMF8688 (R.F.). Z.W. acknowledges funding provided by the Institute for Quantum Information and Matter, an NSF Physics Frontiers Center (NSF Grant PHY-2317110). This research was done using services provided by the OSG Consortium \cite{osg1, osg2, osg3, osg4}, which is supported by the National Science Foundation awards \#2030508 and \#2323298. The minimum-cost flow simulations were performed using Google OR-Tools \cite{ortools}.
% \end{acknowledgments}

\bibliographystyle{apsrev4-2-author-truncate} % Shortens author lists, use for arxiv submission
\bibliography{refs}

\makeatletter 
    
\renewcommand\onecolumngrid{% <<<<<<
\do@columngrid{one}{\@ne}%
\def\set@footnotewidth{\onecolumngrid}% <<<<<<<<<<<<<<<<
\def\footnoterule{\kern-6pt\hrule width 1.5in\kern6pt}%
}

\renewcommand\twocolumngrid{% <<<<<<
        \def\footnoterule{% restore rule
        \dimen@\skip\footins\divide\dimen@\thr@@
        \kern-\dimen@\hrule width.5in\kern\dimen@}
        \do@columngrid{mlt}{\tw@}
}%

\makeatother

\onecolumngrid
\newpage

\renewcommand{\thefigure}{S\arabic{figure}}
\renewcommand{\theequation}{S\arabic{equation}}
\renewcommand{\thetable}{S\Roman{table}}
\renewcommand{\thesection}{S\Roman{section}}

\setcounter{secnumdepth}{2}
\setcounter{equation}{0}
\setcounter{section}{0}
\setcounter{figure}{0}
\setcounter{table}{0}
\setcounter{page}{1}

\renewcommand*{\thefootnote}{\fnsymbol{footnote}}
\setcounter{footnote}{0}
\footnotetext{%
\hypertarget{equalcontrib}{}
\noindent \href{mailto:vlad.temkin@berkeley.edu}{vlad.temkin@berkeley.edu} \\
\noindent \href{mailto:zackmw@caltech.edu}{zackmw@caltech.edu} \\
\noindent V.T. and Z.W. contributed equally to this work.
}
\setcounter{footnote}{0}
\renewcommand*{\thefootnote}{\arabic{footnote}}

\begin{center}
\textbf{\large Supplemental Material For:
Charge-Informed Quantum Error Correction}

\vspace{5mm}
Vlad~Temkin,\textsuperscript{1,\hyperlink{equalcontrib}{\textasteriskcentered}} Zack~Weinstein,\textsuperscript{1,2,\hyperlink{equalcontrib}{\textasteriskcentered}} Ruihua~Fan,\textsuperscript{1} Daniel~Podolsky,\textsuperscript{3}
and Ehud~Altman\textsuperscript{1,4} 

\vspace{1mm}

\textsuperscript{1}\textit{\small Department of Physics, University of California, Berkeley, California 94720, USA} \\
% \vspace{-1mm}
\textsuperscript{2}\textit{\small Department of Physics and Institute for Quantum Information and Matter, \\
\vspace{-0.5mm}
California Institute of Technology, Pasadena, California 91125, USA} \\
\vspace{-0.5mm}
\textsuperscript{3}\textit{\small Physics Department, Technion, Haifa 32000, Israel} \\
\vspace{-0.5mm}
\textsuperscript{4}\textit{\small Materials Sciences Division, Lawrence Berkeley National Laboratory, Berkeley, California 94720, USA} \\
\vspace{-0.5mm}
{\small (Dated: \today)}
\end{center}

\tableofcontents

\section{Statistical Physics Mapping}
\label{app:stat-physMap}
In this Appendix, we discuss the statistical physics model introduced in the main text in detail. We first demonstrate several different equivalent representations of the model, and then we describe how the winding number variance can be interpreted in these models. We also discuss how the R\'enyi-1 and fidelity correlators \cite{weinsteinR12025,lessaSWSSB2025}, two observables for diagnosing strong-to-weak spontaneous symmetry breaking (SWSSB) \cite{leeJY2024SWSSB,lessaSWSSB2025,sala_spontaneous_2024}, can be represented in these models. Finally, we briefly mention some R\'enyi-2 observables which exhibit quantitatively different critical phenomena, but may nevertheless provide useful qualitative intuition on proper R\'enyi-1 observables.

\subsection{Models and Dualities}
As discussed in the main text, we consider an $L \times L$ periodic square lattice of quantum rotors, whose integer-valued quanta represent anyonic quasiparticles carrying U(1) charge. As a minimal model of low-frequency noise applied to our encoded topological state, we consider a decoherence process where, for each nearest-neighbor link $(i\mu)$ of the lattice ($\mu \in \qty{x,y}$), $k_{i\mu}$ quasiparticles are created at site $i$ and $k_{i\mu}$ quasiholes are created at site $i + e_{\mu}$ with a probability $p_{k_{i \mu}} \propto e^{-k_{i \mu}^{2} / 2\alpha}$. Consequently, each total ``error'' in our system is labeled by an integer-valued vector field $k \equiv \qty{k_{i \mu}}$ on the lattice, and the charge distribution $n \equiv \qty{n_{i}}$ created by such an error is given by the lattice divergence of $k$:
\begin{equation}
    n_{i} = \Delta_{\mu} k_{i\mu} \equiv k_{ix} + k_{iy} - k_{i - e_{x}, x} - k_{i - e_{y}, y} .
\end{equation}
Another error configuration $k'$ with the same charge distribution (i.e., $\Delta_{\mu} k_{i \mu}' = n_{i}$) differs from $k$ by a divergenceless vector field $J \equiv \qty{J_{i \mu}}$ satisfying $\Delta_{\mu} J_{i \mu} = 0$. Denoting the probability of the error $k$ as $P_k \equiv \prod_{i\mu} p_{k_{i\mu}}$, the total probability of observing the charge distribution $n = \Delta k$ is obtained by summing over all such divergenceless vector fields:
\begin{equation}
    \label{eq:Zkloops}
    \begin{split}
        Z_{k} &= \sum_{k' : \Delta k' = \Delta k} P_{k'} = \sum_{J : \Delta J = 0} P_{k + J} \\
        &\propto \sum_{J : \Delta J = 0} \exp \qty{ - \frac{1}{2\beta} \sum_{i\mu} (k_{i \mu} + J_{i \mu})^{2} } .
    \end{split}
\end{equation}
Note in particular that $Z_{k} = Z_{k+J}$ is independent of the choice of representative $k$. In the second line above, we have replaced $\alpha$ in the probability distribution $p_{k_{i\mu}}$ with a new parameter $\beta$, so that $Z_{k}$ can be regarded as the partition function of a disordered statistical physics model independent of the error correction problem at hand. $Z_{k}$ corresponds to the probability of observing the charge distribution $n = \Delta k \equiv \qty{\Delta_{\mu} k_{i \mu}}$ upon setting $\beta = \alpha$, which serves as a Nishimori condition on the disordered statistical model \cite{nishimori1981InternalEnergy,nishimoriBook2001}.

In the special case $k = 0$ (i.e., $k_{i \mu} = 0$ for all $i \mu$), $Z_{0}$ can be interpreted as the clean partition function of a standard current-loop model, which is well-known to be dual to the Villain XY model \cite{jose_renormalization_1977,chaikin2013CMbook}. We can derive a similar model in the disordered case by identical techniques. First, we impose the divergenceless constraint at each site $i$ using a Lagrange multiplier $\theta_{i} \in [0, 2\pi)$:
\begin{equation}
    \begin{split}
        Z_{k} &\propto \sum_{J} \int_{\theta} \exp \qty{ - \frac{1}{2\beta} \sum_{i \mu} (k_{i \mu} + J_{i \mu})^{2} - i \sum_{i} \theta_{i} \Delta_{\mu} J_{i \mu} } \\
        &= \sum_{J} \int_{\theta} \exp \qty{ - \frac{1}{2\beta} \sum_{i \mu} J_{i \mu}^{2} - i \sum_{i} \theta_{i} \Delta_{\mu} (J_{i \mu} - k_{i \mu}) } \\
        &= \sum_{J} \int_{\theta} \exp \qty{ - \frac{1}{2\beta} \sum_{i \mu} J_{i \mu}^{2} + i \sum_{i \mu} J_{i \mu} \Delta_{\mu} \theta_{i} + i \sum_{i} n_{i} \theta_{i} } ,
    \end{split}
\end{equation}
where $\int_{\theta} \equiv \int \prod_{i} \frac{\dd{\theta_{i}}}{2\pi}$. In the second line we have shifted $J_{i \mu} \to J_{i \mu} - k_{i \mu}$, while in the third line we have performed a discrete integration by parts and used $\Delta_{\mu} k_{i \mu} = n_{i}$. We have also defined $\Delta_{\mu} \theta_{i} \equiv \theta_{i + e_{\mu}} - \theta_{i}$. As a final step, we employ the Poisson summation formula\footnote{Recall that the Poisson summation formula states that $\sum_{m \in \mathbb{Z}} f(2\pi m) = \frac{1}{2\pi} \sum_{k \in \mathbb{Z}} \tilde{f}(k)$ for a smooth function $f(x)$ and its Fourier transform $\tilde{f}(k) \equiv \int \dd{x} e^{-i k x} f(x)$. In particular, $\sum_{m \in \mathbb{Z}} \exp{- \frac{\beta}{2}(x - 2\pi m)^{2} } = \frac{1}{\sqrt{2\pi \beta}} \sum_{J \in \mathbb{Z} }\exp{- \frac{1}{2\beta}J^{2} + i Jx}$.} to trade $J_{i \mu}$ for an integer-valued gauge field $m_{i \mu}$ on each link of the lattice:
\begin{equation}
    \label{eq:Zk_final}
    Z_{k} \propto \sum_{m} \int_{\theta} \exp \qty{ - \frac{\beta}{2} \sum_{i \mu} (\Delta_{\mu} \theta_{i} - 2\pi m_{i \mu})^{2} + i \sum_{i} n_{i} \theta_{i} } .
\end{equation}
Note that this form of $Z_{k}$ is manifestly independent of the choice of representative $k$.

The case $k = 0$ precisely returns the two-dimensional Villain XY model, which is well-known to exhibit a short-range correlated phase at small $\beta$ and a power-law correlated phase at large $\beta$, separated by a Berezinskii-Kosterlitz-Thouless (BKT) transition \cite{jose_renormalization_1977,jankeVillainMC1991}. Interestingly, the effect of disorder in this representation is to introduce quenched ``clock charges'' $e^{i n_{i} \theta_i }$ at each site, which locally break the U(1) symmetry of the model. Since the total charge is constrained to vanish, the global U(1) symmetry $\theta_{i} \mapsto \theta_{i} + \phi$ is preserved. In the limit of weak disorder, the clock charges are largely organized into dilute dipole pairs, and it is natural to expect that they would have a weak effect on the long-distance physics of the XY model; this intuition is formalized in Appendix~\ref{sec:weak_disorder}. On the other hand, the strong-disorder limit features many far-separated clock charges which locally break the symmetry, despite the global U(1) symmetry in each disorder realization.

\begin{figure}
     \centering
         \includegraphics[width=0.9\textwidth]{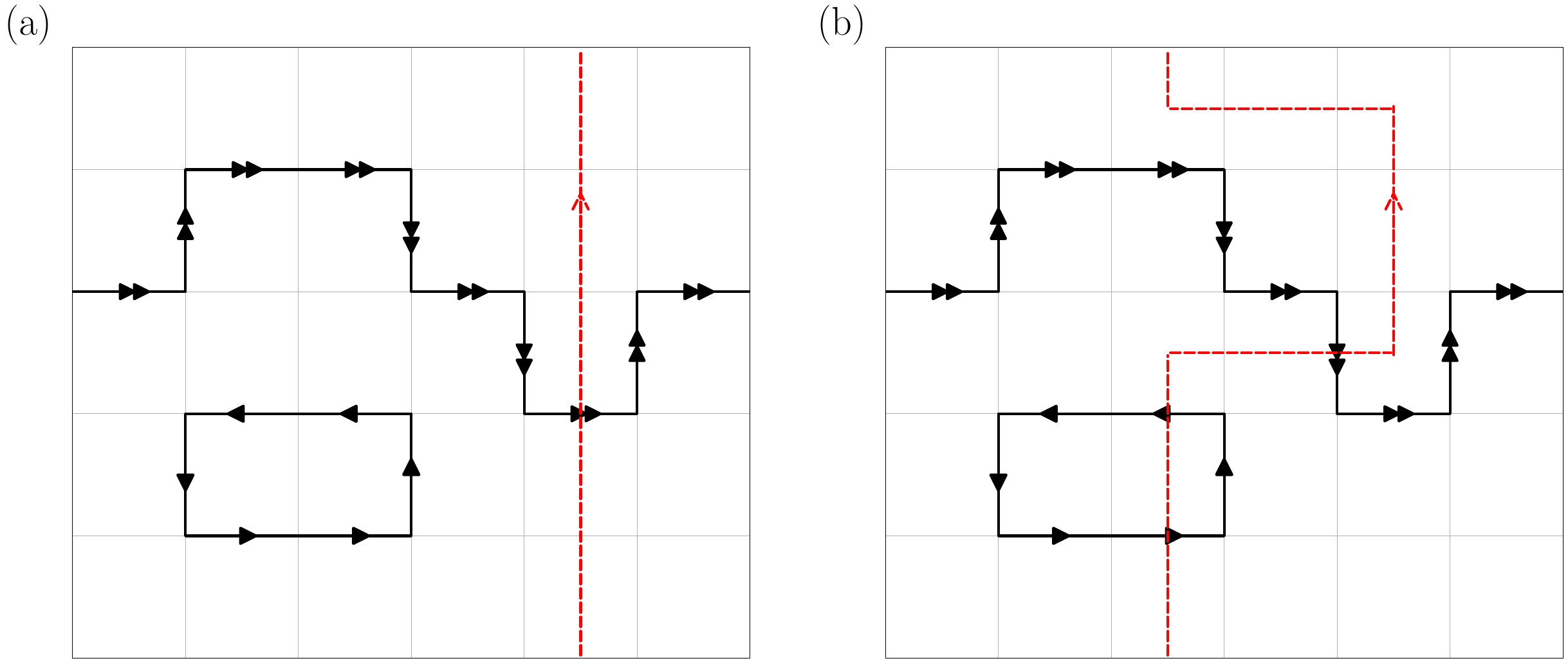}
        \caption{(a) The winding number $W_{x}$ of a current loop configuration $J$ (solid black lines), obtained by summing over all $J_{i\mu}$ cut by the curve $C_{y}$ (dotted red line). In this case, $W_{x} = 2$. (b) The same winding number computed along a different equivalent curve. $W_{x}$ is invariant under sequential deformations of the curve $C_{x}$.}
        \label{fig:cut}
\end{figure}

The partition function $Z_{k}$ contains a sum over all winding sectors of $J$. The winding number $W_{x}$ of $J$ in the $x$ direction can be computed by summing $J_{i \mu}$ along any directed non-contractible loop $C_{y}$ through the dual lattice oriented in the $y$ direction (and analogously for $W_{y}$, using a non-contractible loop $C_{x}$ oriented in the $x$ direction):
\begin{equation}
    W_{x} = \sum_{(i\mu) \in C_{y}} J_{i\mu}, \quad W_{y} = \sum_{(i\mu) \in C_{x}} J_{i \mu} .
\end{equation}
These integer-valued sums are invariant under sequential deformations of the curves $C_{y}$ and $C_{x}$, owing to the divergenceless constraint\footnote{In continuous notation, we are computing $W_{x} = \oint_{C_{y}} \star J$ and $W_{y} = \oint_{C_{x}} \star J$, where $J = J_{\mu}(x) \dd{x^{\mu}}$ is regarded as a one-form satisfying the divergenceless constraint $\dd \star J = \partial^{\mu} J_{\mu} = 0$. Stoke's theorem then implies that $\oint_{C} \star J = 0$ whenever $C = \partial R$ is a homologically trivial closed loop (namely, the boundary of a region $R$). Consequently, we can freely continuously deform $C_{y}$ and $C_{x}$ without changing the value of the winding number.} on $J$ (see Figure \ref{fig:cut}). In particular, we can compute $W_{\mu}$ simply by adding the above equations over $L$ parallel straight-line cycles, giving the simpler formula:
\begin{equation}
\label{eq:winding_2}
    W_{\mu} = \frac{1}{L} \sum_{i} J_{i \mu} ,
\end{equation}
which again takes integer values. Perfect error correction is possible if, for each typical error $k$, the winding number $W_{\mu}$ is zero and nonfluctuating, so that the winding sector of $k$ can be correctly inferred from the observed charge distribution. 

It shall be useful to divide the partition function $Z_{k}$ into contributions from each winding sector:
\begin{equation}
\label{eq:winding_decomposition}
    Z_{k} = \sum_{W_{x}, W_{y} \in \mathbb{Z}} Z_{k}^{(W_{x}, W_{y})}, 
\end{equation}
where the sub-partition function $Z_{k}^{(W_{x}, W_{y})}$ contains only loop configurations with winding numbers $(W_{x}, W_{y})$. In particular, $Z_{k}^{(0, 0)}$ is the total probability of all errors within the same winding sector as $k$. It can be written as the partition function of a disordered height model by expressing $J_{i \mu} = \varepsilon_{\mu \nu} \Delta_{\nu} h_{\tilde{i}}$ as the curl of an integer-valued height field\footnote{As a minor technical subtlety, note that $\Delta_{\mu} h_{\tilde{i}}$ should be interpreted as a \textit{backwards} difference (i.e., $\Delta_{\mu} h_{\tilde{i}} \equiv h_{\tilde{i}} - h_{\tilde{i} - e_{\mu}}$) for $\Delta_{\mu} J_{i \mu} = 0$ to be satisfied. Additionally, the ``zero-mode'' $\sum_{\tilde{i}} h_{\tilde{i}}$ is ill-defined. Since the Boltzmann weight is independent of this zero mode, its inclusion would result in an infinite contribution to the partition function.} $h_{\tilde{i}}$, where $\tilde{i} \equiv (i_{x} + \frac{1}{2}, i_{y} + \frac{1}{2})$ denotes a site on the dual lattice:
\begin{equation}
\label{eq:height_model}
    Z_{k}^{(0, 0)} = \sum_{h} \exp \qty{ - \frac{1}{2\beta} \sum_{i \mu} (\Delta_{\mu} h_{\tilde{i}} - \varepsilon_{\mu \nu} k_{i \nu})^{2} } .
\end{equation}
In this representation, $k_{i \mu}$ enters in the partition function as ordinary bond disorder in the height model. In the same way that a random-bond Ising model can be viewed as containing open-string domain walls which terminate on frustrated plaquettes, the above disordered height model can be viewed as containing ``Escher staircases'' about each dual-lattice plaquette (i.e., each direct lattice site $i$) where $\Delta_{\mu} k_{i \mu}$ is nonzero. 

Finally, it is interesting to represent $Z_{k}$ in the height model representation, and $Z_{k}^{(0, 0)}$ in the XY representation. In the former case, we can \textit{locally} decompose $J_{i \mu} = \varepsilon_{\mu \nu} \Delta_{\nu} h_{\tilde{i}}$ as the curl of a height field, but $h_{\tilde{i}}$ is now multivalued on the torus\footnote{More concretely, we can simply replace $\Delta_{\mu} h_{\tilde{i}}$ in the action with $\Delta_{\mu} h_{\tilde{i}} - \varepsilon_{\mu \nu} W_{\nu}$ along a single row or column of the periodic lattice. However, this is only one of many choices which can be made. A more ``covariant'' option is to write $J_{i \mu} = \varepsilon_{\mu \nu} (\Delta_{\nu} h_{\tilde{i}} - b_{\tilde{i} \nu})$ using an integer-valued gauge field $b_{\tilde{i} \nu}$, which is constrained to exhibit zero flux ($\varepsilon_{\mu \nu} \Delta_{\mu} b_{\tilde{i} \nu} = 0$), but may have nontrivial holonomies $\oint \dd{x}^{\mu} b_{\mu}$ (properly discretized) about the cycles of the torus. The resulting model is invariant under gauge transformations $h_{\tilde{i}} \mapsto h_{\tilde{i}} + \ell_{\tilde{i}}$, $b_{\tilde{i} \mu} \mapsto b_{\tilde{i} \mu} + \Delta_{\mu} \ell_{\tilde{i}}$, and the previous option corresponds to only one possible gauge choice.}. In particular, it satisfies $h_{\tilde{i} + L e_{\mu}} = h_{\tilde{i}} - \varepsilon_{\mu \nu} W_{\nu}$, where $W_{\nu}$ is now regarded as an independent variable to be summed in the partition function. In the latter case, we can enforce the zero-winding constraint with additional Lagrange multipliers $\phi_{x}$ and $\phi_{y}$ by multiplying the Boltzmann weight of $Z_{k}$ by $e^{-i (\phi_{x} W_{x} + \phi_{y} W_{y})}$, and then integrating $\phi_{x}$ and $\phi_{y}$ from zero to $2\pi$. After carrying out the same steps as before, the result is a partition function of the form
\begin{equation}
    \label{eq:Zk00_final}
    Z_{k}^{(0, 0)} \propto \sum_{m} \int_{\theta} \int_{\phi_{x}, \phi_{y}} \exp \qty{ - \frac{\beta}{2} \sum_{i \mu} (\Delta_{\mu} \theta_{i} - a_{i \mu} - 2\pi m_{i \mu})^{2} - i \sum_{i\mu} k_{i \mu} (\Delta_{\mu} \theta_{i} - a_{i \mu})  }, \quad a_{i \mu} = \frac{\phi_{\mu}}{L} .
\end{equation}
The quantities $\phi_{\mu}$ can therefore be interpreted as a U(1) gauge flux threaded through the holes of the torus, and $e^{-i \phi_{\mu} W_{\mu}}$ is the Aharanov-Bohm phase of boson worldlines which encircle the torus. Equivalently, we may perform a gauge transformation to eliminate $a_{i \mu}$, at the cost of imposing the twisted boundary conditions $\theta_{i + L e_{\mu}} = \theta_{i} + \phi_{\mu}$.

\subsection{Helicity Modulus}
As mentioned in the previous section, perfect error correction is possible if and only if the winding number is zero and nonfluctuating; that is, if $\overline{\expval{W_{\mu}^{2}}} \equiv \sum_{k} P_{k} \expval{W_{\mu}^{2}}_{k}$ vanishes. This quantity is closely related to the average helicity modulus $\overline{\Upsilon} \equiv \sum_{k} P_{k} \Upsilon_{k}$, which is defined for each disorder realization by
\begin{equation}
    \Upsilon_{k} \equiv \frac{1}{2} \sum_{\mu = x,y} \qty[ \expval{W_{\mu}^{2}}_{k} - \expval{W_{\mu}}_{k}^{2} ] .
\end{equation}
In clean model with $k = 0$, $\expval{W_{\mu}^{2}}_0$ and $\Upsilon_0$ are equal, and are both directly related to the \textit{phase rigidity} of the XY spins. In this section, we review the relation between $\overline{\Upsilon}$ and the phase rigidity. 

To begin, let us define the partition function $Z_{k}(\phi)$ to be given by Eq.~\eqref{eq:Zk_final}, but with a flux $\phi_{\mu}$ threaded through each $\mu$th hole of the torus. Concretely,
\begin{equation}
\label{eq:Zk_phi}
    Z_{k}(\phi) \propto \sum_{m} \int_{\theta} \exp \qty{ -\frac{\beta}{2} (\Delta_{\mu} \theta_{i} - a_{i \mu} - 2\pi m_{i \mu})^{2} - i \sum_{i \mu} k_{i \mu} (\Delta_{\mu} \theta_{i} - a_{i \mu}) }, \quad a_{i \mu} = \frac{\phi_{\mu}}{L} .
\end{equation}
Recall that a similar construction was previously introduced to represent $Z_{k}^{(0, 0)}$ in the XY representation in Eq.~\eqref{eq:Zk00_final}; in particular, we have $\int_{\phi_{x}, \phi_{y}} Z_{k}(\phi) = Z_{k}^{(0, 0)}$. Performing the same calculation in reverse, we can represent $Z_{k}(\phi)$ in the current-loop representation as
\begin{equation}
    Z_{k}(\phi) = \sum_{J : \Delta J = 0} \exp \qty{ - \frac{1}{2\beta} \sum_{i \mu} (k_{i \mu} + J_{i \mu})^{2} - i \sum_{\mu = x,y} W_{\mu} \phi_{\mu} } ,
\end{equation}
with winding numbers $W_{\mu}$ as defined in Eq.~\eqref{eq:winding_2}. Finally, by differentiating $\log Z_{k}(\phi)$ twice with respect to $\phi_{\mu}$, we arrive at the result
\begin{equation}
    \label{eq:freeEnHelMod}
    \overline{\Upsilon} = - \frac{1}{2} \sum_{\mu = x, y} \pdv[2]{\phi_{\mu}} \overline{\log Z_{k}(\phi)} \eval_{\phi = 0} .
\end{equation}
Altogether, we find that $\overline{\Upsilon}$ is simply related to the disorder-averaged free energy cost of inserting a small flux through a hole of the torus. After a gauge transformation, we can alternatively regard $\overline{\Upsilon}$ as the disorder-averaged free energy cost of introducing the twisted boundary conditions $\theta_{i + L e_{\mu}} = \theta_{i} + \phi_{\mu}$. Such twisted boundary conditions cost vanishingly little free energy in the thermodynamic limit when the XY spins are disordered, but $\mathcal{O}(1)$ free energy in a two-dimensional system when the spins exhibit phase rigidity. A nonzero $\overline{\Upsilon}$ is therefore a direct probe of phase rigidity in the XY spins. In particular, it is well-known that $\Upsilon_0$ in the clean model exhibits a universal jump from zero to $2/\pi$ at the BKT transition into the power-law correlated phase \cite{nelson1977Stiffness,chaikin2013CMbook}.

In the clean XY model, $\expval{W_{\mu}}_{0} = 0$ by symmetry and $\Upsilon_{0} = \expval{W_{\mu}^{2}}_{0}$. In contrast, it is possible in the disordered model for $W_{\mu}$ to be both nonzero and nonfluctuating within a given disorder realization $k$, i.e., $\expval{W_{\mu}^{2}}_{k} = \expval{W_{\mu}}^{2}_{k}$, so that $\overline{\Upsilon}$ is vanishing. This behavior can be probed by the statistical variance of $\expval{W_{\mu}}_k$ over disorder realizations $k$, which we call the \emph{Edwards-Anderson helicity modulus} and denote by $\chi$: 
\begin{equation}
\label{eq:chi}
    \chi \equiv \frac{1}{2} \sum_{\mu = x,y} \qty[ \sum_{k} P_{k} \expval{W_{\mu}}_{k}^{2} - \qty( \sum_{k} P_{k} \expval{W_{\mu}}_{k} )^{2} ] = \frac{1}{2} \sum_{\mu = x, y} \sum_{k} P_{k} \expval{W_{\mu}}_{k}^{2} , 
\end{equation}
where $\overline{\expval{W_{\mu}}} \equiv \sum_{k} P_{k} \expval{W_{\mu}}_{k}$ vanishes by symmetry. We refer to a phase in which $\chi$ is nonvanishing while $\overline{\Upsilon}$ vanishes as a \textit{loop-glass} phase. As we demonstrate in Appendix~\ref{app:constraints}, $\overline{\Upsilon} = \chi$ along the Nishimori line $\alpha = \beta$, and therefore the Nishimori line cannot exhibit a loop-glass phase.

\subsection{R\'enyi-1 and Fidelity Correlators}
Besides the helicity modulus, another fundamental probe of the BKT transition in clean XY models is the two-point correlator $\expval{e^{i(\theta_{x} - \theta_{y})}}_{0}$, which exhibits exponential decay in the high-temperature phase and power-law decay in the low-temperature phase, with a power-law exponent proportional to the inverse helicity modulus \cite{chaikin2013CMbook}. In this section, we will demonstrate that the two-point correlator in the disordered case is directly related to the R\'enyi-1 correlator $R_{1}(x,y)$, defined here as
\begin{equation}
    R_{1}(x,y) \equiv \tr[e^{i(\hat{\varphi}_{x} - \hat{\varphi}_{y})} \sqrt{\rho_{\alpha}} e^{-i(\hat{\varphi}_{x} - \hat{\varphi}_{y})} \sqrt{\rho_{\alpha}}] .
\end{equation}
As discussed in Ref.~\cite{weinsteinR12025}, the R\'enyi-1 correlator provides a natural diagnostic for strong-to-weak spontaneous symmetry breaking (SWSSB), which can be interpreted as a certain conventional symmetry-breaking transition in the canonical purification of $\rho_{\alpha}$. We note that in the present model, the R\'enyi-1 correlator is exactly equal to the fidelity correlator $F(x,y) \equiv \tr \sqrt{ \sqrt{\rho_\alpha} e^{i(\hat{\varphi}_x - \hat{\varphi}_y)} \rho_\alpha e^{-i(\hat{\varphi}_x - \hat{\varphi}_y)} \sqrt{\rho_{\alpha}} }$, \cite{lessaSWSSB2025}, another diagnostic for SWSSB. 

To compute $R_{1}(x,y)$, we shall treat the charged quasiparticles as local excitations and ignore the winding sector of the errors $k$ entirely; that is, we shall take our phenomenological rotor model introduced in the main text literally. Then, since each error $k$ produces the charge distribution $n = \Delta k$, $\rho_{\alpha}$ is given by
\begin{equation}
    \rho_{\alpha} = \sum_{k} P_{k} \dyad{\Delta k} ,
\end{equation}
where $\ket{\Delta k} \equiv \prod_{i} \ket{\Delta_{\mu} k_{i \mu}}$ is a product state in the charge basis with $n_{i} = \Delta_{\mu} k_{i \mu}$ charges on each site $i$. To simplify, we group together all terms which produce the same charge distribution:
\begin{equation}
    \rho_{\alpha} = \sum_{n} \mathcal{Z}_{n} \dyad{n}, \quad \mathcal{Z}_{n} \equiv \sum_{k : \Delta k = n} P_{k} .
\end{equation}
Note that $\mathcal{Z}_{n}$ is the total probability of achieving the charge distribution $n$, and is therefore precisely equal to $Z_{k}$ for any error satisfying $\Delta k = n$. In this form, $\rho_{\alpha}$ is diagonal, and $R_{1}$ can easily be computed; in particular, we can easily compute the square root $\sqrt{\rho_{\alpha}} = \sum_{n} \sqrt{\mathcal{Z}_{n}} \dyad{n}$. 

Let $n^{xy} \equiv \qty{n^{xy}_i}$ be the charge distribution constructed from $n$ by adding one charge at site $x$ and removing one charge at site $y$; i.e.,
\begin{equation}
    n^{xy}_i = \begin{cases}
        n_i + 1, & i = x \\
        n_i, & i \neq x,y \\
        n_i - 1, & i = y
    \end{cases} .
\end{equation}
Then, $R_1(x,y)$ can be conveniently expressed as
\begin{equation}
    \begin{split}
        R_{1}(x,y) &= \sum_{n, n'} \sqrt{ \mathcal{Z}_{n} \mathcal{Z}_{n'}} \tr[e^{i(\hat{\varphi}_{x} - \hat{\varphi}_{y})} \dyad{n} e^{-i(\hat{\varphi}_{x} - \hat{\varphi}_{y})} \dyad{n'} ] = \sum_{n, n'} \sqrt{\mathcal{Z}_{n} \mathcal{Z}_{n'}} \delta_{n', n^{xy}} = \sum_{n} \sqrt{\mathcal{Z}_{n} \mathcal{Z}_{n^{xy}}} \\
        &= \sum_{n} \mathcal{Z}_{n} \sqrt{\frac{\mathcal{Z}_{n^{xy}}}{\mathcal{Z}_{n}}} .
    \end{split}
\end{equation}
We can simplify further by expanding $\mathcal{Z}_{n}$ as a conditional sum over errors $k$. Let $k^{xy}$ be the error obtained from $k$ by adding an additional string of errors from $x$ to $y$, so that $\Delta k^{xy} = n^{xy}$. Then, using $\mathcal{Z}_{n} = Z_{k}$ and $\mathcal{Z}_{n^{xy}} = Z_{k^{xy}}$, we have
\begin{equation}
    R_{1}(x,y) = \sum_{n} \sum_{k : \Delta k = n} P_{k} \sqrt{ \frac{Z_{k^{xy}}}{Z_{k}} } = \sum_{k} P_{k} \sqrt{ \frac{Z_{k^{xy}}}{Z_{k}} } ,
\end{equation}
where we have combined the sum over $n$ and the conditional sum over $k$ into a single unconditional sum on $k$. Finally, using the representation \eqref{eq:Zk_final} of $Z_{k}$, we see that the only difference in $Z_{k^{xy}}$ and $Z_{k}$ is the insertion of an additional factor $e^{i(\theta_{x} - \theta_{y})}$ inside the summation. We therefore arrive at the final expression
\begin{equation}
    \label{eq:R1map}
    R_{1}(x,y) = \sum_{k} P_{k} \sqrt{\expval{e^{i(\theta_{x} - \theta_{y})}}_{k}} .
\end{equation}
In other words, $R_{1}(x,y)$ simply probes the typical behavior of two-point correlations in the disordered statistical physics model along its Nishimori line.

As a final note, we mention that disorder averages of ordinary \textit{linear} correlation functions in the statistical model are trivial, on average:
\begin{equation}
    \sum_{k} P_{k} \expval{ e^{i \theta_{x_{1}}} \ldots e^{i \theta_{x_{p}}} e^{-i \theta_{y_{1}}} \ldots e^{-i \theta_{y_{p}}} }_{k} = 1.
\end{equation}
This is a peculiar feature of our disordered statistical physics model with a Nishimori condition that relates the disorder probability distribution $P_{k}$ with the Boltzmann weights in the partition function\footnote{In the Nishimori random-bond Ising model (with a sum over periodic and antiperiodic boundary conditions on the torus), the analogous statement is that the average behavior of disorder parameter correlations is trivial.}. The proof of this relation is essentially the reverse of the preceding derivation:
\begin{equation}
    \sum_{k} P_{k} \expval{ e^{i \theta_{x_{1}}} \ldots e^{i \theta_{x_{p}}} e^{-i \theta_{y_{1}}} \ldots e^{-i \theta_{y_{p}}} }_{k} = \sum_{k} P_{k} \frac{Z_{\tilde{k}}}{Z_{k}} = \sum_{n} \sum_{k : \Delta k = n} P_{k} \frac{Z_{\tilde{k}}}{Z_{k}} = \sum_{n} \mathcal{Z}_{n} \frac{\mathcal{Z}_{\tilde{n}}}{\mathcal{Z}_{n}} = \sum_{n} \mathcal{Z}_{\tilde{n}} = \sum_{n} \mathcal{Z}_{n} = 1.
\end{equation}
Here $\tilde{k}$ is a modified error which inserts additional charges at $x_{1},\ldots,x_{p}$ and removes charges at $y_{1},\ldots,y_{p}$, and $\tilde{n} = \Delta \tilde{k}$. In the final expression, we have used the fact that $\mathcal{Z}_{n}$ is the probability of observing the charge distribution $n$, and the sum over all such probabilities must be unity.

\subsection{R\'enyi-2 Observables}
Although the capacity for performing error correction is most directly related to the ``R\'enyi-1'' quantities defined above, useful qualitative information can often be drawn from simpler R\'enyi-2 quantities \cite{fanDiagnostics2024,bao_mixedstate_2023,leeJY2024SWSSB,sala_spontaneous_2024}. Whereas the R\'enyi-1 quantities introduced above are quenched-average observables in a disordered system, R\'enyi-2 observables are obtained by performing an \emph{annealed} average over the disorder.

Concretely, we study the R\'enyi-2 correlator $R_{2}(x,y)$, defined by
\begin{equation}
    R_{2}(x,y) \equiv \frac{\tr[e^{i(\hat{\varphi}_{x} - \hat{\varphi}_{y})} \rho_{\alpha} e^{-i(\hat{\varphi}_{x} - \hat{\varphi}_{y})} \rho_{\alpha}]}{\tr \rho_{\alpha}^{2}} .
\end{equation}
Following a nearly identical calculation to the previous section, $R_{2}(x,y)$ is reduced to
\begin{equation}
    R_{2}(x,y) = \frac{\sum_{n} \mathcal{Z}_{n} \mathcal{Z}_{n^{xy}}}{\sum_{n} \mathcal{Z}_{n}^{2}} = \frac{\sum_{k} P_{k} Z_{k^{xy}}}{\sum_{k} P_{k} Z_{k}} ,
\end{equation}
which shows that $R_{2}$ is precisely the annealed average of the correlation function $\expval{e^{i(\theta_{x} - \theta_{y})}}_k$. The annealed average of the partition function in the denominator is given by
\begin{equation}
    \begin{split}
        \sum_{k} P_{k} Z_{k} &\propto \sum_{k} e^{-\frac{1}{2\alpha} \sum_{i \mu} k_{i \mu}^{2} } \sum_{m} \int_{\theta} \exp \qty{ - \frac{\alpha}{2} (\Delta_{\mu} \theta_{i} - 2\pi m_{i \mu})^{2} - i \sum_{i \mu} k_{i \mu} \Delta_{\mu} \theta_{i} } \\
        &\propto \sum_{m, m'} \int_{\theta} \exp \qty{ - \frac{\alpha}{2} (\Delta_{\mu} \theta_{i} - 2\pi m_{i \mu})^{2} - \frac{\alpha}{2} \sum_{i \mu} (\Delta_{\mu} \theta_{i} - 2\pi m'_{i \mu})^{2} } ,
    \end{split}
\end{equation}
where we have performed the sum over $k$ using the Poisson summation formula, introducing a \textit{second} integer gauge field $m'_{i \mu}$. We see that R\'enyi-2 correlations in our model are described by a Villain XY model with two distinct integer gauge fields. At large $\alpha$, we expect the above annealed partition function to be dominated by configurations with zero vortices, and we effectively recover a Villain XY model at inverse temperature $2\alpha$. This suggests that $R_{2}(x,y)$ decays as a power law in the large-$\alpha$ regime, and that the transition from exponential decay to power-law decay is governed by an ordinary BKT transition at a critical threshold $\alpha_{c}^{(2)}$.

We can also analyze the helicity modulus of the annealed-average partition function, defined by
\begin{equation}
    \begin{split}
        \Upsilon^{(2)} &\equiv - \frac{1}{2} \sum_{\mu = x,y} \pdv[2]{\phi_{\mu}} \log \qty{ \sum_{k} P_{k} Z_{k}(\phi) } \eval_{\phi = 0} \\
        &= \frac{1}{2 \sum_{k} P_{k} Z_{k}} \sum_{\mu = x, y} \qty[ \sum_{k} P_{k} \sum_{J : \Delta J = 0} P_{k+J} W_{\mu}^{2} ] ,
    \end{split}
\end{equation}
where $Z_{k}(\phi)$ was defined in Eq.~\eqref{eq:Zk_phi}. Due to the simple structure of the annealed average partition function, it is natural to expect that $\Upsilon^{(2)}$ exhibits a universal jump of $2/\pi$ at the critical point $\alpha_{c}^{(2)}$.

\section{Optimal and Suboptimal Decoders}
\label{app:opt_subopt}
In this Appendix, we review the precise sense in which the disordered statistical physics model described above with the Nishimori condition $\alpha = \beta$ constitutes an ``optimal'' error correction scheme. We will also describe how the models with $\beta \neq \alpha$ can be viewed as alternative suboptimal decoders, and how the loop-glass phase of our statistical physics model should be interpreted through the lens of error correction.

To emphasize the dependence of the partition functions $Z_{k}$ [Eq.~\eqref{eq:Zkloops}] and the fixed-winding partition functions $Z_{k}^{(W_{x}, W_{y})}$ [Eq.~\eqref{eq:winding_decomposition}] on the temperature $\beta$, we shall write $Z_{k}(\beta)$ and $Z_{k}^{(W_{x}, W_{y})}(\beta)$ in this section. This notation should not be confused with $Z_{k}(\phi)$ in Eq.~\eqref{eq:Zk_phi}, where $\phi$ denoted a flux threaded through the holes of the torus.

Starting from an encoded state with no quasiparticles, we suppose that an unknown error $\hat{E}_k$ occurs with probability $P_k$. We postulate that the global charge distribution $n = \Delta k$ can be measured, and we are tasked with determining a recovery operator $\hat{E}_r^{\dag}$ which recovers the initial encoded state. Such a recovery will successfully eliminate all quasiparticles as long as $\Delta r = \Delta k$, i.e., the vector field $J = k - r$ is divergenceless. We further postulate that $\hat{E}_r^{\dag}$ correctly recovers the encoded state if $J$ has zero winding, i.e., if $J_{i\mu} = \varepsilon_{\mu \nu} \Delta_{\mu} h_{\tilde{i}\nu}$ is the curl of a height field on the dual lattice. On the other hand, recoveries which induce a nonzero winding number will generally induce a logical error on the encoded state. 

The optimal decoder (i.e., the optimal error correction scheme) is the algorithm for choosing $r$ which recovers the original encoded state with the highest possible probability, given the information available. Given an observed syndrome $n$, the quantity $Z_{k}^{(W_x, W_y)}(\alpha)$ is proportional to the \textit{conditional} probability of all errors in the winding sector $(W_x, W_y)$ (defined relative to $k$), conditioned on the syndrome $n$. Therefore, the conditional probability that a recovery $r$ with winding $(W_x, W_y)$ successfully recovers the original encoded state is simply
\begin{equation}
    \text{Prob} \Big( k \in \text{Winding Sector }(W_{x}, W_{y}) \  \Big\vert \ n \Big) = \frac{Z_{k}^{(W_{x}, W_{y})}(\alpha)}{Z_{k}(\alpha)} .
\end{equation}
The optimal decoding strategy\footnote{Strictly speaking, if the quasiparticles created by $\hat{E}_k$ are anyons of order $m$ (i.e., the fusion of $m$ such anyons yields a trivial particle), then windings $W_{\mu} \in m \mathbb{Z}$ do not lead to a logical error. The true optimal decoder therefore computes the probabilities of the error classes $\sum_{\ell, \ell' \in \mathbb{Z}} Z_k^{(w + \ell m, w' + \ell' m)}$ for $w, w' = 0, \ldots , m-1$, where $Z_k^{(W_x, W_y)}$ is the probability of the winding sector $(W_x, W_y)$. Our physical assumption is that higher-order windings are substantially suppressed compared to small windings $W_{\mu} = 0, \pm 1$, so that this truly optimal decoder is essentially equivalent to correcting in the single most likely winding sector. With this subtlety in mind, we shall continue to refer to our decoder as the optimal decoder.} is to compute $Z_k^{(W_x, W_y)}(\alpha)$ for each winding sector, and choose a recovery $r$ in the winding sector with the largest probability. Note that the observer does not know the \textit{absolute} values of the winding numbers in advance, since these are defined relative to the unknown error $k$. Instead, the observer can choose an arbitrary error $k^{n}$ consistent with $n$ and define winding numbers relative to $k^{n}$.

For $\beta \neq \alpha$, the quantity $Z_{k}^{(W_{x} , W_{y})}(\beta)$ is not directly related to the probability of the winding sector $(W_{x}, W_{y})$. Instead, for each $\beta$, we can construct a suboptimal decoder from these partition functions by interpreting $Z_{k}^{(W_{x}, W_{y})}(\beta)$ as a ``confidence'' in the winding sector $(W_{x}, W_{y})$. Given the syndrome $n$, this decoder computes these confidences for each winding sector, and applies a recovery in the most ``confident'' sector. We can regard decoders with $\beta < \alpha$ as biased towards more likely individual errors, while decoders with $\beta > \alpha$ can be regarded as biased towards winding sectors with a larger number of probable errors. In particular, the decoder with $\beta = 0$ places all of its confidence in the winding sector with the single most likely error; this is the \textit{minimum-cost flow} decoder, discussed in the main text and in Appendix~\ref{app:min_cost_flow}. Only the optimal decoder with $\alpha = \beta$ correctly balances the relative importance of ``energy'' (i.e., the probability of individual errors) and ``entropy'' (i.e., the number of errors in each winding sector).

In a given disorder realization $k$, a decoder with zero winding number variance (i.e., $\Upsilon_{k}(\beta) = 0$) places all of its confidence in a single winding sector, while a decoder with $\Upsilon_{k}(\beta) \neq 0$ is not perfectly confident in its guess at the winding of $k$. Since these confidences coincide with the actual probabilities of each winding sector when $\alpha = \beta$, we expect that the optimal decoder succeeds in decoding for typical $k$ when $\Upsilon_{k}(\alpha) = 0$, and fails with a nonzero probability when $\Upsilon_{k}(\alpha) \neq 0$. We shall formalize this intuition in Appendix~\ref{app:constraints}, where we prove that $\overline{\Upsilon} = \chi$ on the Nishimori line; that is, whenever the optimal decoder is completely confident for typical errors, it also correctly predicts the winding of these typical errors.

On the other hand, for $\beta \neq \alpha$, it is possible that $\Upsilon_{k} = 0$ but $\expval{W_{\mu}}_{k} \neq 0$. In this case, the suboptimal decoder is ``confidently incorrect'': it places all of its confidence in the wrong winding sector. This scenario is precisely what occurs in the loop-glass phase of our model, where $\overline{\Upsilon} = 0$ while $\chi \neq 0$: with nonzero probability, the decoder places all of its confidence in a nonzero winding sector, leading to $\overline{\expval{W_{\mu}}^{2}} > 0$.

\section{Constraints on the Phase Diagram from the Nishimori Condition}
\label{app:constraints}
In Appendix~\ref{app:opt_subopt}, we described how the model \eqref{eq:Zkloops} can be understood as defining a large class of decoders for the U(1) symmetric noise model discussed in the main text. We have also discussed how the average mean-square winding number, $\overline{\expval{W_{\mu}^2}}$, is zero and nonfluctuating whenever the decoder succeeds with high probability, and is nonzero when the decoder fails. Since the decoder with $\beta = \alpha$ is the optimal decoder, we expect that no other decoder can possibly exhibit zero mean-square winding number for disorder strengths beyond the Nishimori critical point $\alpha_c$; that is, for all $\alpha > \alpha_c$, the model is necessarily in either a large-loop phase with $\overline{\Upsilon} > 0$ or a loop-glass phase with $\overline{\Upsilon} = 0$ and $\chi > 0$. Furthermore, we expect that the optimal decoder can never be ``confidently incorrect'' in its decoding, so that the Nishimori line cannot exhibit a loop-glass phase.

Here we shall prove the above two statements, utilizing the special structure of the model imposed by the Nishimori condition. The calculations presented in this section follow simply and immediately from nearly identical calculations in the $\pm J$ random-bond Ising model (RBIM) \cite{nishimoriBook2001}, and have previously been used to place similar constraints on that model's phase diagram.

Our derivation will crucially rely on the \textit{gauge structure} of the partition functions $Z_k$, which depend only on the charge distribution $n = \Delta k$ of the disorder realization $k$. Explicitly, $Z_{k+J} = Z_k$ for any divergenceless current $J$, and any observable $O[J]$ satisfies the following simple constraint:
\begin{equation}
    \expval{O[J]}_{k+J'} \equiv \frac{1}{Z_{k+J'}} \sum_{J : \Delta J = 0} P_{k+J'+J} O[J] = \frac{1}{Z_k} \sum_{J : \Delta J = 0} P_{k+J} O[J - J'] = \expval{O[J-J']}_k ,
\end{equation}
where $P_{k + J} \propto \exp \qty{ - \frac{1}{2\beta} \sum_{i\mu} (k_{i\mu} + J_{i\mu})^2 }$. Note that the above holds both on the Nishimori line and away from it. 

To derive general constraints on the phase diagram, let us consider the squared winding number $W^2_{\mu}$ as our observable, i.e., $O[J] = W^2_{\mu}[J] = (\frac{1}{L} \sum_i J_{i\mu})^2$. The above calculation gives
\begin{equation}
\label{eq:gauge_identity}
    \expval{W_{\mu}^2}_{k + J'} = \expval{W_{\mu}^2}_k + W_{\mu}^2[J'] - 2 W_{\mu}[J'] \expval{W_{\mu}}_k .
\end{equation}
To utilize this equation, let us consider the disorder average $\overline{\expval{W_{\mu}^2}^{(\beta)}}$ at temperature $\beta$. Since this includes a sum over all disorder realizations $k$, we can freely average $\expval{W_{\mu}^2}^{(\beta)}_k$ over all gauge transformations of $k$ without affecting the result. This leads to the following interesting sequence of equalities:
\begin{equation}
    \begin{split}
        \overline{\expval{W_{\mu}^2}^{(\beta)}} &\equiv \sum_k P_k^{(\alpha)} \expval{W_{\mu}^2}^{(\beta)}_k \\
        &= \frac{1}{N_J} \sum_{J' : \Delta J' = 0} \sum_{k} P_{k+J'}^{(\alpha)} \expval{W_{\mu}^2}^{(\beta)}_{k+J'} \\
        &= \frac{1}{N_J} \sum_{J' : \Delta J' = 0} \sum_k P^{(\alpha)}_{k + J'} \qty[ \expval{W^2_{\mu}}^{(\beta)}_k + W^2_{\mu}[J'] - 2 W_{\mu}[J'] \expval{W_{\mu}}_k^{(\beta)} ] \\
        &= \frac{1}{N_J} \sum_k Z_k^{(\alpha)} \qty[ \expval{W_{\mu}^2}^{(\beta)}_k + \expval{W_{\mu}^2}^{(\alpha)}_k - 2 \expval{W_{\mu}}^{(\alpha)}_k \expval{W_{\mu}}^{(\beta)}_k ] .
    \end{split}
\end{equation}
The second line utiilzes the average over all $N_J$ gauge transformations\footnote{Strictly speaking, the number $N_J$ of gauge transformations is infinite. We expect that this should not pose any real problem to the derivation, since the probabilities $P_{k+J}^{(\alpha)}$ are exponentially decaying for large $J$, and so we can always restrict the sum over gauge transformations to a large but finite set.}, the third line uses the identity \eqref{eq:gauge_identity}, and the final line reinterprets the disorder average as a thermal average at temperature $\alpha$. Finally, we can write $Z_k^{(\alpha)} = \sum_{J : \Delta J = 0} P_{k + J}^{(\alpha)}$ as a sum over probabilities, and interpret the sum over $J$ as another sum over gauge transformations:
\begin{equation}
    \begin{split}
        \overline{\expval{W_{\mu}^2}^{(\beta)}} &= \frac{1}{N_J} \sum_k \sum_{J: \Delta J = 0} P_{k+J}^{(\alpha)} \qty[ \expval{W_{\mu}^2}^{(\beta)}_k + \expval{W_{\mu}^2}^{(\alpha)}_k - 2 \expval{W_{\mu}}^{(\alpha)}_k \expval{W_{\mu}}^{(\beta)}_k ] \\
        &= \sum_k P_k^{(\alpha)} \qty[ \expval{W_{\mu}^2}^{(\beta)}_k + \expval{W_{\mu}^2}^{(\alpha)}_k - 2 \expval{W_{\mu}}^{(\alpha)}_k \expval{W_{\mu}}^{(\beta)}_k ] \\
        &= \overline{\expval{W^2_{\mu}}^{(\beta)}} + \overline{\expval{W^2_{\mu}}^{(\alpha)}} - 2 \overline{\expval{W_{\mu}}^{(\alpha)} \expval{W_{\mu}}^{(\beta)}} ,
    \end{split}
\end{equation}
where we have used the fact that the quantity in brackets is invariant under gauge transformations.

Altogether, by comparing the lefthand and righthand sides, we have arrived at the following simple result:
\begin{equation}
    \overline{\expval{W^2_{\mu}}^{(\alpha)}} = 2 \overline{\expval{W_{\mu}}^{(\alpha)} \expval{W_{\mu}}^{(\beta)}} .
\end{equation}
Or equivalently, by setting $\beta = \alpha$ on the right hand side,
\begin{equation}
    \overline{\expval{W_{\mu}}^{(\alpha)} \expval{W_{\mu}}^{(\beta)}} = \overline{\expval{W_{\mu}}^{(\alpha)} \expval{W_{\mu}}^{(\alpha)}} .
\end{equation}

From these results, we can learn several features of the phase diagram. The first immediate result is obtained by setting $\beta = \alpha$, from which we find that the disorder-averaged helicity modulus $\overline{\Upsilon}$ and the Edwards-Anderson helicity modulus $\chi$ are equal along the Nishimori line:
\begin{equation}
    \label{eq:NLrelation}
    \overline{\Upsilon} = \chi \quad (\alpha = \beta),
\end{equation}
which implies that the Nishimori line cannot exhibit a loop-glass phase with $\overline{\Upsilon} = 0$ and $\chi > 0$. 

Another immediate result is obtained by setting $\beta = 0$, where we find that if $\expval{W_{\mu}}_k^{(\beta = 0)}$ vanishes for all typical disorder realizations $k$, then we must also have $\overline{\expval{W^2_{\mu}}^{(\alpha)}} = 0$. Since $\expval{W_{\mu}}_k^{(\beta = 0)}$ becomes nonzero for typical $k$ past the minimum-cost flow threshold $\alpha_c^{\text{MCF}}$, this implies that
\begin{equation}
    \alpha_c^{\text{MCF}} < \alpha_c ,
\end{equation}
where $\alpha_c$ is the critical point on the Nishimori line. We find as expected that the minimum-cost flow decoder has a lower threshold than the optimal decoder. More generally, along any constant-$\beta$ cut of the phase diagram, $\expval{W_{\mu}}_k^{(\beta)} = 0$ for typical $k$ implies that $\overline{\expval{W_{\mu}^2}^{(\alpha)}_k} = 0$. This implies that no disordered phase can exist beyond $\alpha_c$, i.e., any point in the phase diagram for $\alpha > \alpha_c$ must be either in the long-loop phase or a loop-glass phase.

\section{Weak Disorder Field Theory}
\label{sec:weak_disorder}

In this Appendix, we discuss an effective field-theoretic treatment of the model \eqref{eq:Zk_final} in the limit of low temperatures\footnote{Note that $\beta \gg 1$ is the limit of low temperature for the Villain XY model, but high temperature for the loop model.} $\beta \gg 1$ and weak disorder $\alpha \ll 1$. From the field theory we develop, we shall demonstrate that the phase transition between the short-loop and large-loop phases remains BKT-like for sufficiently weak disorder, albeit with a modified and non-universal jump in the helicity modulus.

We start from the XY representation of the partition function $Z_k$, given in Eq.~\eqref{eq:Zk_final}. Writing the disorder explicitly in terms of $k_{i\mu}$ rather than its divergence $n_i$, the partition function is given by
\begin{equation}
        Z_{k} \propto \sum_{m} \int_{\theta} \exp \qty{ - \frac{\beta}{2} \sum_{i \mu} (\Delta_{\mu} \theta_{i} - 2\pi m_{i \mu})^{2} - i \sum_{i\mu} k_{i\mu} \Delta_\mu \theta_{i} } .
\end{equation}
Recall that each $k_{i\mu}$ is an integer-valued random variable, sampled with probability $p_{k_{i\mu}} \propto e^{-k_{i\mu}^2/2\alpha}$. In the clean limit $\alpha = 0$, $Z_k$ is the partition function of the ordinary Villain XY model.

\subsection{Field theory}
In the low-temperature phase $\beta > \beta_c^{\rm BKT}$, the XY model is described at long distances by a field theory of a compact boson $\theta(x)$ with the action
\begin{equation}
    \mathcal{S}[\theta] = \frac{K}{2} \int \dd[2]{x} (\partial_{\mu} \theta)^{2} ,
\end{equation}
where $K$ is the helicity modulus. Since $\Delta_{\mu} \theta_i \simeq \partial_{\mu} \theta(x_i)$ is slowly varying on the lattice scale, we can include disorder by coarse-graining the term $\sum_i k_{i\mu} \Delta_{\mu} \theta_i$. Specifically, we can sum $k_{i\mu}$ over a volume $B(i\mu)$ where $\Delta_{\mu} \theta_i$ remains approximately constant. When $V = \abs{B(i\mu)}$ becomes large, the central limit theorem allows us to regard the resulting coarse-grained disorder $k_{\mu}(x_i) \simeq \frac{1}{V} \sum_{i'\mu' \in B(i\mu)} k_{i'\mu'}$ as a Gaussian random variable with zero mean and variance
\begin{equation}
    \label{eq:WDvar}
    \expval{k^{\mu}(x) k^{\nu}(y)} = D \delta^{\mu \nu} \delta(x-y), \qquad D \simeq \frac{\sum_{k \in \mathbb{Z}} k^2 e^{-k^2/2\alpha}}{\sum_{k \in \mathbb{Z}} e^{-k^2/2\alpha}} =  \frac{\qty( -\partial_{(2\alpha)^{-1}}) \theta_3\qty(0, e^{-1/2\alpha})}{\theta_3\qty(0, e^{-1/2\alpha})},
\end{equation}
where $\theta_3\qty(0, e^{-1/2\alpha})$ is the Jacobi theta function. We therefore obtain the following disordered theory in the continuum:
\begin{equation}
    \label{eq:disTheoryNoVortices}
    \mathcal{S}[\theta] = \int \dd[2]{x} \qty{ \frac{K}{2} (\partial_{\mu} \theta)^{2} + i k^{\mu} \partial_{\mu} \theta }.
\end{equation}

For $\beta \gg \beta_c^{\text{BKT}}$, we can neglect vortices entirely and straightforwardly use the above field theory to determine the effect of disorder on the helicity modulus. Recall from Eq.~\eqref{eq:freeEnHelMod} that the disorder-averaged helicity modulus can be obtained from the average free energy cost of imposing twisted boundary conditions $\theta_{i + L e_{\mu}} = \theta_{i} + \phi_{\mu}$. Here we shall perform the disorder average using the replica trick \cite{kardarFields2007,nishimoriBook2001}. In the present context, it is convenient to formulate the replica trick using the following identity:
\begin{equation}
    \overline{\log \mathcal{Z}_k} = \lim_{R \to 0} \frac{1}{R} \log \overline{\mathcal{Z}_k^{R}} ,
\end{equation}
where $\mathcal{Z}_k \equiv \int \mathcal{D}\theta \, e^{-\mathcal{S}[\theta]}$. Introducing $R$ replicas, and averaging over disorder, we obtain the following replicated partition function:
\begin{equation}
\label{eq:replicaTheory}
    \begin{split}
        \overline{\mathcal{Z}_k^R} &= \int \prod_{r = 1}^R \mathcal{D} \theta_r \, e^{-\mathcal{S}_R[\theta_r]}, \\
        \mathcal{S}_{R}[\theta_{r}] &= \int \dd[2]{x} \qty{ \frac{K}{2} \sum_{r = 1}^{R} (\partial_{\mu} \theta_{r})^{2} + \frac{D}{2} \sum_{r, r' = 1}^{R} (\partial_{\mu} \theta_{r}) (\partial^{\mu} \theta_{r'})}.
    \end{split}
\end{equation}

To obtain the helicity modulus, we impose the twisted boundary conditions $\theta(x + L e_{\mu}) = \theta(x) + \phi_{\mu}$. Since $\mathcal{S}_{R}$ is Gaussian, we can compute the path integral directly via the saddle-point method \cite{chaikin2013CMbook}. Specifically, we set
\begin{equation}
    \theta(x) = \tilde{\theta}(x) + \sum_{\mu = x, y} \frac{\phi_{\mu}}{L} x^{\mu},
\end{equation}
where $\tilde{\theta}(x)$ is a smooth periodic function. By a straightforward calculation, we obtain
\begin{equation}
    \log \overline{[\mathcal{Z}_k(\phi)]^R} = \log \overline{\mathcal{Z}_k^R} - \sum_{\mu = x, y} \qty[ R \frac{K}{2} \phi_{\mu}^2 + R^2 \frac{D}{2} \phi_{\mu}^2  ],
\end{equation}
so that the replica limit yields
\begin{equation}
    \label{eq:helModFreeBosonReplicated}
    \overline{\Upsilon} = - \frac{1}{2} \sum_{\mu = x, y} \pdv[2]{\phi_{\mu}} \qty[ \lim_{R \to 0} \frac{1}{R} \log \overline{[\mathcal{Z}_k(\phi)]^R} ] = K.
\end{equation}
We therefore find that the disorder doesn't affect the disorder-averaged helicity modulus for very large $\beta$.

We can also use the replica field theory to compute the Edwards-Anderson helicity modulus $\chi$ [see Eq.~\eqref{eq:chi}]. Towards this end, we first express $\overline{\expval{W_{\mu}}^2}$ in the replicated theory:
\begin{equation}
    \begin{split}
        \overline{\expval{W_{\mu}}^2} &= - \overline{ \qty( \frac{1}{\mathcal{Z}_k} \pdv{\mathcal{Z}_k}{\phi_{\mu}} )^2 } \eval_{\phi = 0} = - \lim_{R \to 0} \frac{1}{\overline{\mathcal{Z}_k^R}} \overline{ \mathcal{Z}_k^{R - 2} \pdv{\mathcal{Z}_k}{\phi_{\mu}} \pdv{\mathcal{Z}_k}{\phi_{\mu}} } \eval_{\phi = 0} \\
        &= - \pdv{\phi_{\mu, 1}} \pdv{\phi_{\mu, 2}} \log \overline{[\mathcal{Z}_k(\phi_1) \ldots \mathcal{Z}_k(\phi_R)]} ,
    \end{split}
\end{equation}
where we have introduced an independent set of twists $\phi_r = (\phi_{x, r}, \phi_{y, r})$ for each $r$th replica. We therefore see that $\overline{\expval{W_{\mu}}^2}$ is related to the free energy cost of twisting two distinct replicas independently, rather than twisting each replica uniformly. By a similar calculation as above, we can again compute the disorder average of this replica free energy:
\begin{equation}
    \log \overline{[\mathcal{Z}_k(\phi_1) \ldots \mathcal{Z}_k(\phi_R)]} = \log \overline{\mathcal{Z}_k^R} - \sum_{\mu = x, y} \qty[ \frac{K}{2} \sum_{r = 1}^R \phi_{\mu, r}^2 + \frac{D}{2} \sum_{r, r' = 1}^R \phi_{\mu, r} \phi_{\mu, r'} ] .
\end{equation}
Taking derivatives with respect to $\phi_1$ and $\phi_2$ and then sending $R \to 0$, we obtain
\begin{equation}
    \chi = \frac{1}{2} \sum_{\mu = x, y} \overline{\expval{W_{\mu}}^2} = D .
\end{equation}

\subsection{Universal and Non-Universal Stiffness Jumps}
To determine the behavior of $\overline{\Upsilon}$ and $\chi$ in the vicinity of the transition, it is necessary to take vortices into account. Towards this end, we employ standard duality techniques to represent the XY model as a Sine-Gordon model \cite{jose_renormalization_1977}, where the cosine creates and annihilates vortices. As we will see, the presence of disorder does not significantly affect the duality transformation. 

Starting from the theory \eqref{eq:disTheoryNoVortices}, we first perform a Hubbard-Stratonovich transformation to obtain the action:
\begin{equation}
    \mathcal{S}\qty[J, \theta] = \int d^2 x \qty{\frac{1}{2 K} \qty(J_\mu + k_\mu)^2 + iJ^\mu\partial_\mu \theta},
\end{equation}
where we have introduced the Hubbard-Stratonovich field $J$, and we have subsequently shifted $J^\mu \to J^\mu + k^\mu$. Next, we separate $\theta(x) = \theta_{s}(x) + \theta_{v}(x)$ into a spin-wave part and a vortex part; $\theta_{s}(x)$ is a smooth single-valued function, while $\theta_{v}(x)$ is singular at the vortex cores. In particular, $\theta_v$ satisfies $\varepsilon^{\mu \nu} \partial_{\mu} \partial_{\nu} \theta_{v}(x) = 2\pi n_{v}(x)$, where $n_{v}(x)$ is the vortex density. Integration over $\theta_{s}$ imposes $\partial_{\mu} J^{\mu} = 0$, which we solve by writing $J^{\mu} = \frac{1}{2\pi} \varepsilon^{\mu \nu} \partial_{\nu} \varphi$. Integrating the last term by parts, we obtain
\begin{equation}
    \mathcal{S}[\varphi, n_{v}] = \int \dd[2]{x} \qty{ \frac{1}{2(2\pi)^{2} K} (\partial_{\mu} \varphi - 2\pi \varepsilon_{\mu \nu}k^\nu)^{2} + i \varphi(x) n_{v}(x)}.
\end{equation}
If we write $n_{v}(x) = \sum_{i = 1}^{N_{v}} n_{i} \delta(x - x_{i})$ for the vortex density, we see that the partition function contains an insertion of $e^{i n_{i} \varphi(x_{i})}$ for each vortex at position $x_{i}$ with vorticity $n_{i}$; thus, the operator $e^{i \varphi(x)}$ can be thought of as the operator which inserts a vortex in $\theta$ at position $x$. 

We now follow the standard strategy of keeping only vortices with vorticity $n_i = \pm 1$. This allows for us to represent the sum over vortex insertions as
\begin{equation}
    \sum_{N = 0}^{\infty} \frac{y^N}{N!} \int d^2x_1 \ldots d^2 x_N\sum_{\qty{n_i = \pm 1}} e^{i n_1 \varphi(x_1) + \ldots + i n_N \varphi(x_N)} = \exp \qty{ 2y \int d^2x \cos \varphi(x) } .
\end{equation}
Higher order vortices will lead to higher harmonics $\cos(p \varphi)$, which are less relevant than $\cos \varphi$. Altogether, we obtain the following disordered sine-Gordon-like action:
\begin{equation}
    \label{eq:disThVort}
    \mathcal{S}[\varphi] = \int \dd[2]{x} \qty{ \frac{1}{2(2\pi)^{2} K} (\partial_{\mu} \varphi - 2\pi \varepsilon_{\mu \nu}k^\nu)^{2} - 2y \cos \varphi } .
\end{equation}
This action can be regarded as the natural continuum formulation of the height model introduced in Eq.~\eqref{eq:height_model}. Analogous to the lattice formulation, the disorder $k^{\mu}(x)$ now enters as a type of bond disorder.

Similar to the previous section, we can now analyze this disordered Sine-Gordon model via the replica trick. Introducing $R$ replicas and averaging over quenched disorder, we obtain
\begin{equation}
\label{eq:disThVortRepl}
    \mathcal{S}_R[\varphi_r] = \int \dd[2]{x} \qty{ \frac{1}{2(2\pi)^2 K} \sum_{r, r' = 1}^R g_{r r'} (\partial_{\mu} \varphi_r) (\partial^{\mu} \varphi_{r'}) - 2 y \sum_{r = 1}^R \cos \varphi_r }, \quad g_{r r'} = \delta_{r r'} - \frac{D}{K + RD} ,
\end{equation}
where we expanded the square in \eqref{eq:disThVort} and used the term proportional to $R k^2$ to rescale $D \to \frac{D K}{K + RD}$ in the disorder probability. The action \eqref{eq:disThVortRepl} is now suitable for the standard techniques, such as momentum-shell renormalization group (RG).

To determine the effect of disorder on the BKT transition, it is sufficient to determine its effect on the RG flow of the fugacity $y$. This will determine the helicity modulus where vortices become relevant and destroy the power-law ordered phase. One approach is to perform momentum-shell RG, perturbatively integrating Fourier modes between $\Lambda$ and $\Lambda - \dd{\ell}$ to determine the effect on $y$. In the present context, we can take the simpler (and equivalent) approach of reading off the scaling dimension of $\cos \varphi_{r}$ from its two-point correlation function. By diagonalizing the quadratic part of the action, we obtain the following intra-replica correlation functions in the limit $y = 0$:
\begin{equation}
    \expval{ e^{i [\varphi_r(x) - \varphi_r(x')]} }_{y = 0} = e^{-\frac{1}{2} \expval{[\varphi_r(x) - \varphi_r(x')]^2}} \sim \frac{1}{\abs{x - x'}^{2\pi K  [g^{-1}]_{r r}}} ,
\end{equation}
where $[g^{-1}]_{r r} = 1 + \frac{D}{K}$ is the diagonal element from the inverse of the matrix $g_{r r'}$. Taking the replica limit, we obtain
\begin{equation}
    \overline{\expval{e^{i [\varphi(x) - \varphi(x')]}}_{y = 0}} \sim \frac{1}{\abs{x - x'}^{2\Delta}}, \quad \Delta = \pi (K + D) .
\end{equation}
We therefore obtain the following RG flow equation for the fugacity, to leading order:
\begin{equation}
    \dv{y}{\ell} = \qty[ 2 - \pi (K + D) ] y + \mathcal{O}(y^{2}) .
\end{equation}
This suggests that the critical phase is \emph{stabilized} by disorder; in particular, the helicity modulus at the transition is reduced from $2 / \pi$ in the clean model to
\begin{equation}
    \label{eq:KcWeakDis}
    \overline{\Upsilon}(\beta_c) = K_{c} = \frac{2}{\pi} - D .
\end{equation}
We therefore find that weak disorder preserves the BKT transition of the model, but with a reduced and nonuniversal jump in the helicity modulus $\overline{\Upsilon}$. On the other hand, since the Edwards-Anderson helicity modulus $\chi$ is given by $D$ in the low-temperature phase, we find that the average variance of the winding number $\overline{\expval{W^2}} \equiv \frac{1}{2} \sum_{\mu} \overline{\expval{W_{\mu}^2}}$ continues to undergo a universal jump of $2/\pi$:
\begin{equation}
    \label{eq:uniWindJump}
    \overline{\expval{W^2}}(\beta_c) = \overline{\Upsilon}(\beta_c) + \chi(\beta_c) = K_c + D = \frac{2}{\pi} .
\end{equation}

Finally, although we initially derived the replica field theory \eqref{eq:replicaTheory} in the weak disorder limit, it is natural to postulate that it describes the entire long-loop phase with renormalized parameters $K$ and $D$. In particular, if this theory describes the Nishimori line $\alpha = \beta$, then the exact relation $\eqref{eq:NLrelation}$ implies that $K = D$. This gives a stiffness jump of
\begin{equation}
    \overline{\Upsilon}(\alpha_c) = K_c = \frac{1}{\pi} \quad (\alpha = \beta) .
\end{equation}

\section{Worm Algorithm}
\label{app:WA}
Here, we explain the idea behind the Worm Algorithm (WA) \cite{prokofievWorm2001}, which we employ for numerical simulations of the model \eqref{eq:Zkloops} at finite temperatures. The partition function of interest is of the following form:
\begin{equation}
    Z = \sum_{J: \Delta J = 0} \exp \qty{ -\sum_{i\mu}V_{i\mu}\qty( J_{i\mu})},
\end{equation}
where $J: \Delta J = 0$ stands for closed-loop configurations, and $V_{i\mu}\qty( J_{i\mu})$ is a local energy cost function associated with the current variable $J_{i\mu}$ on the links $(i \mu)$ of a $d$-dimensional hypercubic lattice. Partition functions of this form naturally arise in the high-temperature expansion of various XY-like statistical physics models \cite{jose_renormalization_1977,prokofievWorm2001}. In our specific case, the function $V_{i\mu}(J_{i \mu}) = \frac{1}{2\beta} (J_{i \mu} + k_{i \mu})^{2}$ incorporates quenched disorder, arising from the initial errors $k_{i \mu}$, which can be sampled i.i.d. on each link with probabilities $p_{k_{i \mu}} \propto e^{-k_{i \mu}^{2} / 2\alpha}$ at the start of the simulation.

The WA's core idea is to update a closed-loop configuration $J$ by temporarily extending the configuration space to include open-loop configurations, in which the zero-divergence condition $\Delta J = 0$ is violated at two points, the ``head'' and the ``tail'' of the ``worm''. This allows for local updates generated by a random walk of the worm's head, otherwise forbidden by the closed-loop constraint. A closed-loop configuration is reached when the random walk returns to its origin, i.e., when the head and the tail of the worm meet. The probabilities of each move correspond to the Boltzmann weights of the segments of the worm, and thus, this procedure satisfies detailed balance. 

In more detail, we begin the simulation initially with the configuration $J_{i \mu} = 0$ everywhere. In each subsequent step, given a closed-loop configuration of $J$, a single Monte Carlo step of the WA proceeds as follows:  
\begin{enumerate}
    \item With uniform probability $p_{\rm site} = \frac{1}{L^d}$, a lattice site $i$ is chosen at random. The worm's head and tail are both initialized at this site.
    \item A direction $\mu = \qty{\pm x_1, \pm x_2, \dots, \pm x_d}$ is chosen at random with uniform probability $p_{\text{dir}} = \frac{1}{2d}$.
    \item With probability 
    \begin{equation}
        p_{\rm move} = \min\qty{1, \frac{e^{-V_{i\mu}\qty( J^\prime_{i\mu})}}{e^{-V_{i\mu}\qty(J_{i\mu})}}} ,
    \end{equation} 
    the worm's head attempts to move from site $i$ to site $i + e_\mu$. If the move is accepted, the head's position is updated, and $J_{i \mu}$ is updated to $J_{i \mu} \pm 1$, where $\pm 1$ is chosen to match the sign of $\mu$. If the move is rejected, the worm's head remains at the same site. 
    \item If the head and tail's positions coincide, end the loop. Otherwise, return to step 2.
\end{enumerate}
Although the divergenceless constraint is violated at the head and tail of the worm during intermediate steps of the loop, each full Monte Carlo step ends with the head meeting the tail, and thus $\Delta J = 0$ is satisfied after each full step. Only such closed-loop configurations contribute to thermodynamic averages of quantities such as the helicity modulus. 

Intermediate configurations of the WA consist of open paths with two unmatched endpoints (the head and the tail of the worm). Although these configurations do not contribute to the partition function, they can be utilized to compute two-point correlation functions. Specifically, the the correlator $\expval{e^{i(\theta_{x} - \theta_{y})}}$ is represented within the high-temperature expansion as a sum over loop configurations $J_{i \mu}$ with divergence $+1$ at site $x$ and divergence $-1$ at site $y$:
\begin{equation}
    \expval{e^{i (\theta_{x} - \theta_{y})}} = \frac{1}{Z} \sum_{J} \delta \big[ \Delta J_{x} = 1, \ \Delta J_{y} = -1, \ \Delta J_{i \neq x, y} = 0 \big] \exp \qty{ - \sum_{i\mu} V_{i \mu}(J_{i \mu}) } .
\end{equation}
Consequently, in the extended configuration space of the worm algorithm, the numerator of $\expval{e^{i (\theta_{x} - \theta_{y})}}$ can be estimated from the relative frequency of intermediate worm states with the head at site $x$ and tail at site $y$. The partition function is similarly estimated by the total frequency of closed loop completions, and the ratio of the two frequencies provides an estimate for the correlation function.

\section{Minimum-Cost Flow}
\label{app:min_cost_flow}
The Minimum Cost Flow (MCF) problem is a fundamental optimization problem on graphs that arises in a wide range of applications, including transportation planning, economic network optimization \cite{ahujaNetwork1993}, and, as discussed in the main text, error correction. It is defined on a directed graph with given ``supply'' and ``demand'' values at the vertices of the graph, and the goal is to transport flow through the network at minimum total cost while satisfying capacity and flow conservation constraints. Polynomial-time algorithms are currently known only for linear and quadratic cost functions, e.g., \cite{orlinMCF1993, laszloQMCF2016}. In this Appendix, we define the MCF problem and discuss our numerical implementation of the MCF decoder introduced in the main text.

\subsection{Overview}
Formally, the problem is defined by the following data:
\begin{enumerate}
    \item A directed graph $G = \qty( V, A)$, where $V$ is the set of vertices and $A$ is the set of directed edges.
    \item A non-negative capacity $u_{ij} > 0$, $\forall \qty( i, j) \in A$.
    \item A supply/demand $b_i$, $\forall i \in V$, s.t. $\sum_i b_i = 0$.
    \item A cost function $C_{ij} \equiv C_{ij}\qty( f_{ij})$ associated with each directed edge $(i,j)$, where $f_{ij}$ denotes the flow on the edge $(i,j)$. The classical MCF assumes a linear cost function $C_{ij}\qty( f_{ij}) = c_{ij} f_{ij}$, with constant coefficients $c_{ij}$. However, we will need to extend this definition to include a quadratic cost function. We will discuss the differences in what follows.
\end{enumerate}
The optimization problem is to find a flow $f$ which minimizes the quantity $\sum_{(i, j) \in A} C_{ij}(f_{ij})$, subject to the two constraints
\begin{equation}
    \begin{split}
        & 0 \leq f_{ij} \leq u_{ij}, \hspace{2.95cm} \forall (i, j) \in A , \\
        & \sum_{k : (i, k) \in A} f_{ik} - \sum_{k : (k, i) \in A} f_{ki} = b_i, \quad  \forall i \in V .
    \end{split}
\end{equation}
Graphically, we can regard $f$ as an integer-valued current which can flow from $i$ to $j$ for each $(i, j) \in A$; this current is conserved at each vertex, aside from the ``sources'' at sites $i$ with $b_i > 0$, and the ``sinks'' at sites $i$ with $b_i < 0$.

For application to the present work, we shall focus on the integer-valued flow problem where $f_{ij} \in \mathbb{Z}, ~\forall \qty( i, j) \in A$. While integer constraints typically make linear programming problems NP-hard, the constraint matrix in the MCF problem is totally unimodular (i.e., the determinant of every square submatrix is $\pm 1$ or $0$). This property ensures that strongly polynomial-time algorithms exist for the integer-valued linear-cost case. Moreover, when the cost function is separable and convex (as in the quadratic case discussed below), the problem can be mapped onto a standard linear MCF problem on an expanded graph \cite{ahujaNetwork1993}. Below we discuss linear and quadratic cases separately. 

\begin{figure}
     \centering
         \includegraphics[width=0.9\textwidth]{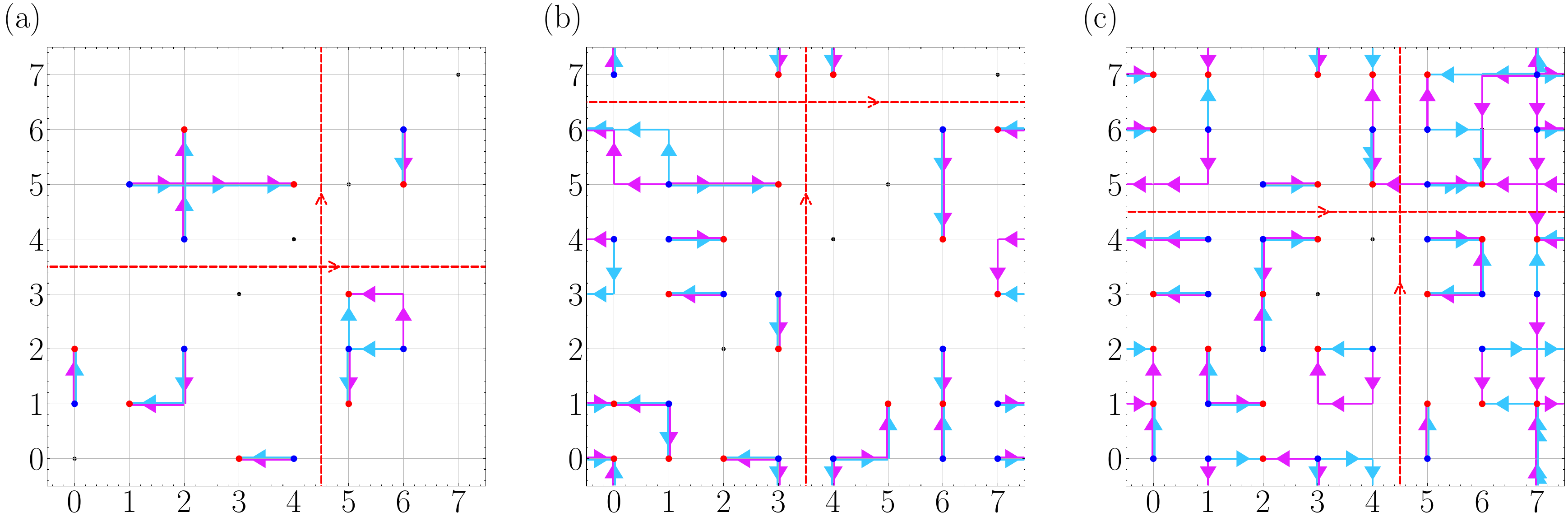}
        \caption{Examples of the MCF solutions for (a) $\alpha = 0.15$, (b) $\alpha = 0.25$, (c) $\alpha = 0.35$ in an $L = 8$ system. Magenta arrows represent the sampled errors, the cyan arrows are the linear MCF solutions. The positive charges are blue, the negative are red. The winding number can be computed along the red dashed cuts. The correction succeeds in (a) and (b) but fails in (c) due to non-trivial winding.}
        \label{fig:MCFconf}
\end{figure}

\begin{enumerate}
    \item {\it Linear Cost Function} \\
    The simplest and most common version of the MCF problem assumes a linear cost function:
    \begin{equation}
        C_{ij}(f_{ij}) = c_{ij} f_{ij}.
    \end{equation}
    Several pseudo-polynomial and strongly-polynomial algorithms have been developed, including Refs.~\cite{orlinMCF1993, ahujaMCF1992, goldbergMCF1997}. In our numerical simulations of the linear MCF, we use the Google OR-Tools library \cite{ortools}, which implements an algorithm based on Ref.~\cite{ahujaMCF1992}.
    \item {\it Quadratic Cost Function} \\
    In some applications, including our zero-temperature decoding procedure, the cost function is quadratic:
    \begin{equation}
        C_{ij}(f_{ij}) = c_{ij} f_{ij}^2.
    \end{equation} 
    A straightforward way to handle this Quadratic Minimum Cost Flow (QMCF) problem is to approximate the quadratic cost by a piecewise linear function. This transforms it into a standard linear MCF problem on an expanded graph \cite{ahujaNetwork1993}. The algorithms for this formulation of the problem are typically pseudo-polynomial, with complexity depending on the edge capacities. Alternatively, there is a strongly polynomial algorithm (i.e., it depends only on the number of vertices and edges of the graph) recently developed in Ref.~\cite{laszloQMCF2016}. However, implementation of these algorithms is numerically challenging. Luckily, we are interested in a range of parameters such that the linear MCF is a very good approximation to the QMCF, see the discussion below. 
\end{enumerate}

\subsection{Linear MCF as an approximation to the quadratic MCF}

As discussed in Appendix~\ref{app:constraints}, the error threshold of the MCF decoder must be lower than that of the optimal decoder, which was found to be $\alpha_c \approx 0.37$. Thus, in our MCF simulations, we are interested in the range of $\alpha$'s below $\alpha_c$.

The quadratic and linear cost functions are identical if the flows $f_{ij}$ take values $0$ or $1$. This is almost always the case when the syndrome (supplies/demands) consists of the $\pm 1$-valued charges, which happens for sufficiently small $\alpha$'s. For a given $\alpha$, the probability of sampling an error valued 2 or greater is
\begin{equation}
p_{|k|\geq 2} = 1 - p_0 - 2p_1 = 1 - \frac{1 + 2 e^{-\frac{1}{2\alpha}}}{\theta_3\qty( 0, e^{-\frac{1}{2\alpha}})}.
\end{equation}
For $\alpha = 0.37$, the probability $p_{|k|\geq 2} \approx 0.006$ is very small. Thus, the ``heavy'' errors occur very rarely, and are well separated from the rest of the errors and easy to correct, so the solutions of the linear and quadratic MCFs are almost always identical in the considered regime of $\alpha$'s.

\subsection{Simulation details}

\begin{figure}
     \centering
         \includegraphics[width=0.6\textwidth]{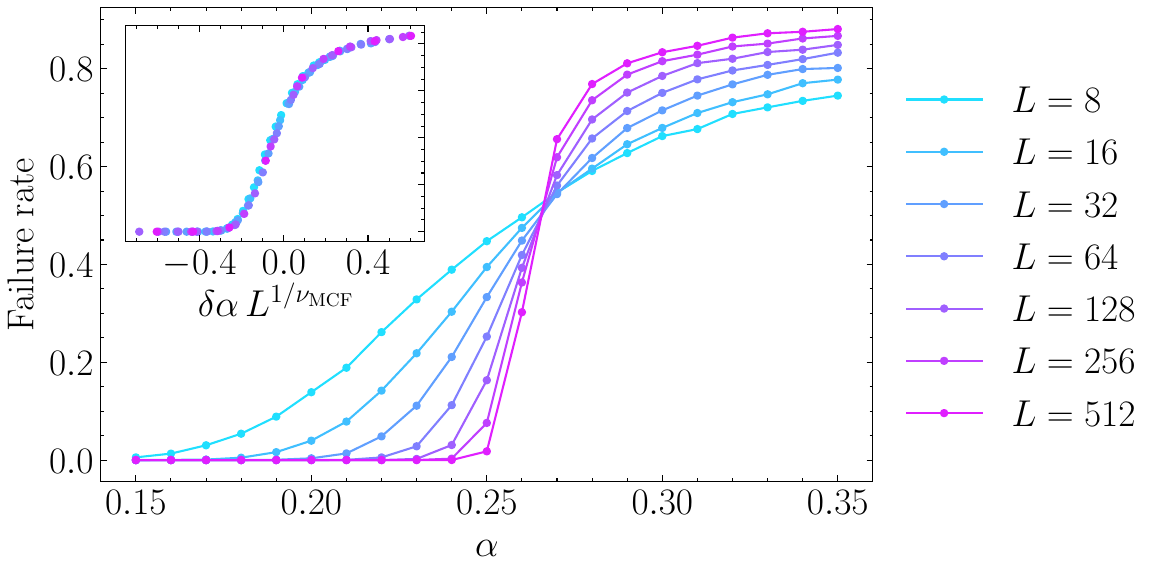}
        \caption{Failure rate of the minimum-cost flow decoder, for system sizes up to $L=512$. Inset: finite-size scaling collapse with $\nu_{\rm MCF} = 2.19$ and $\delta \alpha = \alpha - \alpha_{\text{MCF}}$ with $\alpha_{\text{MCF}} = 0.265$.}
        \label{fig:MCFW}
\end{figure}

As discussed above, we use linear MCF as an approximation to the quadratic MCF. First, we independently sample an error $k_{i\mu}$ with probability $p_{k_{i\mu}} \propto e^{-k_{i\mu}^2/2\alpha}$ on each link of the lattice. Given an error configuration, we determine the syndrome by computing the lattice divergence on each site $n_i = \Delta_\mu k_{i\mu}$. The undirected lattice is further represented as a directed graph $G = (V,A)$, where every link of the lattice is associated with \emph{two} edges directed in the opposite direction between the same two sites. The syndrome charges $n_i$ of the lattice sites are assigned to the supplies/demands $b_i$ on the corresponding graph vertices. The edge capacities are set to infinity $c_{ij} = \infty$ and the cost function is chosen to be $C_{ij} = f_{ij}$ for every edge in the graph. The optimal flow $f_{ij}$ is further computed using the Google-OR \cite{ortools} tools library. The success of the decoding is determined by computing the winding number of the relative loop between the error and correction $(k - f)$. If the winding number is 0, the instance is counted as successful; otherwise, it is counted as a failure. We perform the simulation for $2\cdot10^4$ different disorder configurations for every $\alpha$. The mean-square winding number [see Fig.~2(d) of the main text] and the failure rate (see Fig.~\ref{fig:MCFW}) demonstrate a sharp crossing and an excellent scaling collapse with a finite correlation length exponent $\nu_{\rm MCF} \approx 2.19$.

\section{BKT Finite-Size Scaling}
Here we briefly review the theory of finite-size scaling in BKT-like transitions. We assume RG equations in the vicinity of the transition of the form
\begin{equation}
    \dv{y}{\ell} = \pi (K_c - K) y, \quad \dv{K^{-1}}{\ell} = C y^2 ,
\end{equation}
where $C$ is a nonuniversal constant. In the traditional BKT transition we have $K_c = 2/\pi$, whereas here we would like to allow for a more general critical point. 

It is convenient to simplify these equations by defining $\tilde{x} = \pi K_c^2 (K^{-1} - K_c^{-1})$ and $\tilde{y} = \sqrt{\pi K_c^2 C} y$, after which the RG equations become
\begin{equation}
    \dv{\tilde{x}}{\ell} = \tilde{y}^2, \quad \dv{\tilde{y}}{\ell} = \tilde{x} \tilde{y}.
\end{equation}
Note that the quantity $\tilde{x}^2 - \tilde{y}^2$ is conserved under RG flow, and therefore the flow lines form hyperbolas in $(\tilde{x}, \tilde{y})$ parameter space. The BKT transition is controlled by the separatrix $\tilde{y} = - \tilde{x}$; curves below this line flow to $\tilde{y}(\ell = \infty) = 0$ and $\tilde{x}(\ell = \infty) < 0$ (i.e., $K > K_c$), while curves above this line flow off towards infinity (i.e., $K = 0$). We can therefore set $\tilde{y} = - \tilde{x}$ directly at the critical point, and immediately solve the above differential equation for $\tilde{x}(\ell)$:
\begin{equation}
    \tilde{x}(\ell) = - \frac{1}{\abs{\tilde{x}_0}^{-1} + \ell} ,
\end{equation}
where $\tilde{x}_0 \equiv \tilde{x}(\ell = 0)$ is a negative and non-universal constant. Note the extremely slow flow of $\tilde{x}(\ell)$ towards its fixed-point value $\tilde{x}(\ell) = 0$. Expressing this equation in terms of $K$,
\begin{equation}
    K(\ell) = K_c \qty( 1 + \frac{(\pi K_c)^{-1}}{c + \ell} ) ,
\end{equation}
where $c$ is another non-universal constant.

To derive a finite-size scaling ansatz for the helicity modulus, we regard $\ell$ as a rescaling of the lattice spacing $a$ to $a e^{\ell}$. In a finite-sized system of linear size $L$, the maximum value of $\ell$ is therefore $\log (L/a)$. Since the helicity modulus is controlled by the asymptotic behavior of $K$, we obtain the following finite-size scaling ansatz for $\Upsilon$ as a function of $L$, directly at the critical point:
\begin{equation}
    \label{eq:stiffness_FSSapp}
    \Upsilon(L) = \Upsilon_{\infty} \qty( 1 + \frac{(\pi \Upsilon_{\infty})^{-1}}{ c + \log L} ) ,
\end{equation}
where $\Upsilon_{\infty} = K_c$ is the value of the helicity modulus at the critical point in an infinite system, i.e., the value of the stiffness jump.

Away from the critical point in the disordered phase, by integrating the RG equations for $\tilde{x}^2 - \tilde{y}^2 < 0$, the correlation length $\xi$ can be shown to diverge as $\exp{ a [T - T_c]^{-1/2} }$ \cite{chaikin2013CMbook,kardarFields2007} with another nonuniversal constant $a$. Therefore, a natural finite-size scaling ansatz from the disordered phase is given by \cite{sandvik_computational_2010}
\begin{equation}
    \Upsilon(T, L) = \Upsilon_{\infty} \qty( 1 + \frac{(\pi \Upsilon_{\infty})^{-1}}{ c + \log L} ) \Phi(Le^{-a[T - T_c]^{-1/2}}) ,
\end{equation}
where $\Phi$ is a universal scaling function. We have used this scaling ansatz in the main text for our finite-size scaling collapse of $\overline{\Upsilon}$.

\section{Additional Numerical Results}
Here, we provide additional numerical details for the results from the main text and discuss the behavior of the three-dimensional model on the NL.

\subsection{Outline of the numerics}
\begin{figure}
     \centering
         \includegraphics[width=0.4\textwidth]{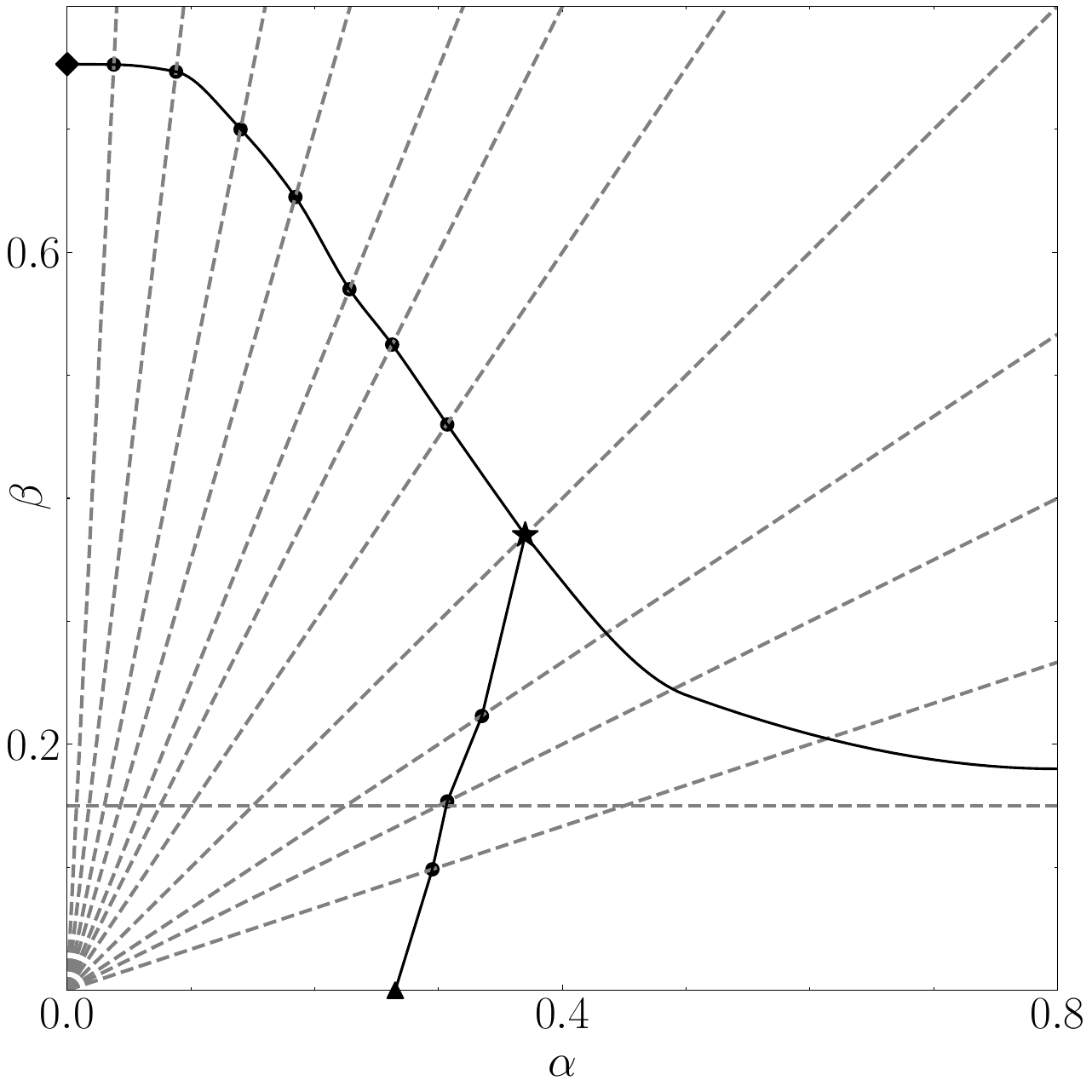}
        \caption{Cuts of the phase diagram used to determine the phase boundaries and properties of the phases. The precision below the NL is limited due to the numerical hardness of that regime.}
        \label{fig:cuts}
\end{figure}

We study the phase diagram of the loop model \eqref{eq:Zkloops}, both on and away from the Nishimori line. We compute the disorder-averaged helicity modulus $\overline{\Upsilon}$ and the Edwards-Anderson helicity modulus $\chi$ along several cuts of the form $\beta = \lambda \alpha$ with $\lambda = \left\{ 20, \frac{17}{2}, 5, \frac{7}{2}, \frac{5}{2}, 2, \frac{3}{2}, 1, \frac{2}{3}, \frac{1}{2}, \frac{1}{3}\right\}$ and a cut with fixed $\beta = 0.15$ (see Fig.~\ref{fig:cuts}). On the Nishimori line, we average over $2000$ disorder samples far from the critical point, $16000$ samples in the vicinity of the critical point for $L < 128$, and $10000$ samples in the vicinity of the critical point for $L=128$. For $\lambda = \left\{ \frac{5}{2}, 2, \frac{3}{2}\right\}$, we sample $1000$ disorder realizations away from the critical point, and $2000$ close to the critical point for every system size. We generate $500$ disorder samples for $\lambda = \left\{ \frac{17}{2}, 5, \frac{7}{2}\right\}$, and $400$ for $\lambda = 20$. Below the Nishimori line, the dynamics are extremely slow. For $\lambda = \frac{2}{3}$, we average over $2000$ samples for $L = 8, 16, 32$, $1000$ samples for $L = 64$, and $500$ samples for $L=128$. For $\lambda = \frac{1}{2}$, we average over 2000 disorder realizations for $L<64$, and 1000 realizations for $L=64$. For $\lambda = \frac{1}{3}$, we average over $1000$ samples for $L < 32$ and $500$ for $L=32$. We sample $1000$ disorder realizations along the $\beta = 0.15$ line for every system size. 

During the thermalization step for $\lambda > 1$, we sample $2.5 \cdot 10^6$ closed loops to ensure thermalization (this is much more than needed in the critical phase and deep in the disordered phase of the clean model). For $\lambda < 1$, we find that in most of the considered cases the system's energy thermalizes after less than $500$ closed loops are sampled. The error bars everywhere are estimated as $\sigma/\sqrt{N}$, where $\sigma$ is the single-sample standard deviation computed with the sampled realizations and $N$ is the number of disorder samples.

Below, we discuss the procedure for estimating the phase boundaries of the short-loop phase. 

\subsection{Nishimori line and weak disorder}

We determine the critical point on the Nishimori line and in the weak disorder limit ($\lambda > 1$) using the procedure developed in Ref.~\cite{weberMCXY1988}. At the BKT transition in the disordered model, the helicity modulus obeys the scaling ansatz \eqref{eq:stiffness_FSSapp}. The critical point is then determined by numerically fitting $\overline{\Upsilon}(\alpha, L)$ to this ansatz for each $\alpha$ and choosing $\alpha_c$ as the minimum of the root mean square (RMS) error. 

The procedure works due to the following reason. The scaling ansatz \eqref{eq:stiffness_FSSapp} captures the numerical data trivially in two limits: deep in the disordered phase and deep in the critical phase. In the former case, the helicity modulus evaluated in finite systems is 0, and equation~\eqref{eq:stiffness_FSSapp} perfectly describes this behavior if $\Upsilon_\infty = 0$ and $c = \infty$ are chosen. In the latter, the helicity modulus in any system size approaches its thermodynamic limit. Thus, the scaling ansatz captures this perfectly by setting $c = \infty$ and $\Upsilon_\infty$ to the numerically obtained value. The RMS of the fit in these two cases is vanishingly small. However, RMS grows in the vicinity of the critical point, and then again gets suppressed exactly at the critical point, exhibiting a local minimum at $\alpha_c$. The RMS error exhibits a sharp minimum in the clean XY model even when very small system sizes are used \cite{weberMCXY1988}. However, in the disordered case, we observe a noisy minimum that can't be used for a systematic critical point determination.

\begin{figure}
     \centering
         \includegraphics[width=\textwidth]{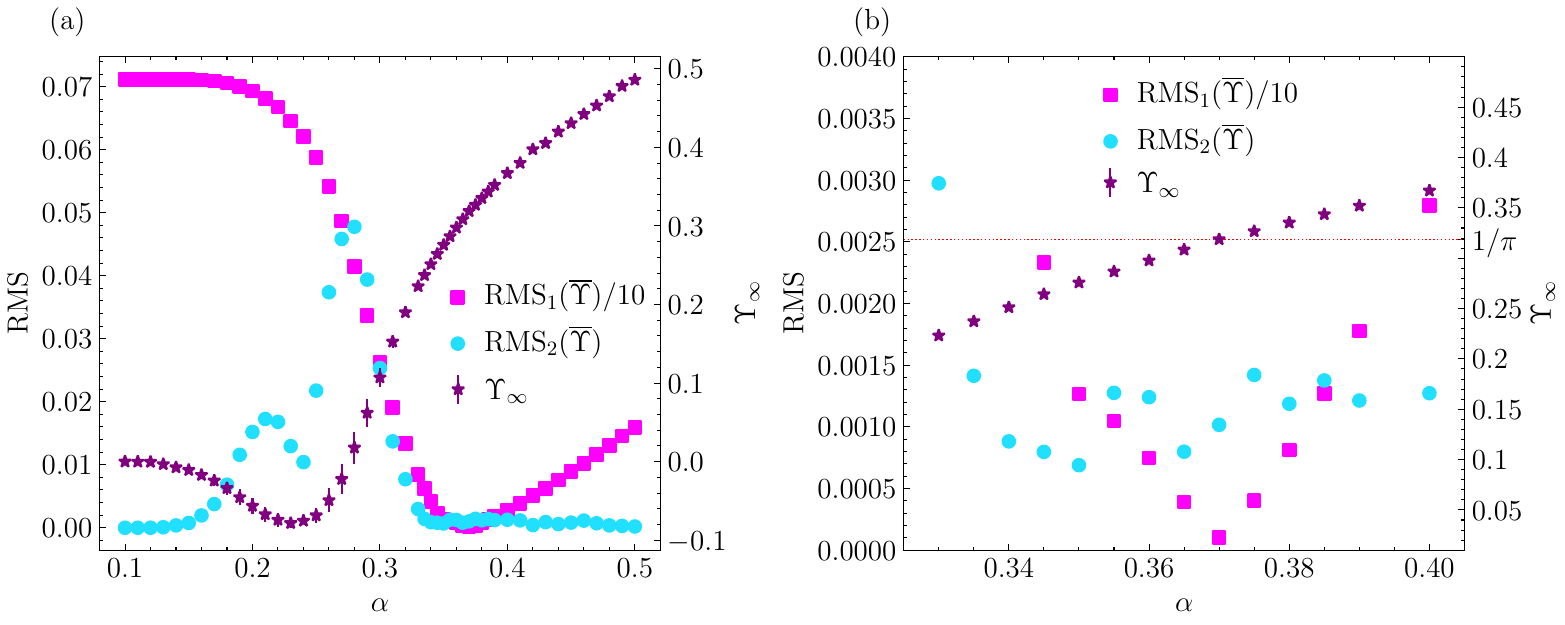}
        \caption{Weber-Minhagen procedure \cite{weberMCXY1988} for extracting the critical point and stiffness jump, applied to the optimal decoder on the Nishimori line $\alpha = \beta$. (a) Blue dots: RMS error of the fits for the scaling collapses \eqref{eq:stiffness_FSSapp} with 2 free parameters. Magenta squares: RMS error of the fits for the scaling collapses \eqref{eq:stiffness_FSSapp} with fixed $\Upsilon_\infty = 1/\pi$ and 1 free parameter, divided by 10. Purple stars: value of the helicity modulus jump $\Upsilon_\infty$, from the two-parameter fit. (b) Close up of (a) around the critical point. We observe a sharp minimum in the winding RMS and a noisy minimum of the helicity modulus RMS. From this plot, we estimate the critical decoherence strength to be $\alpha_c \approx 0.370 \pm 0.005$.}
        \label{fig:RMSNL}
\end{figure}

\begin{figure}
     \centering
         \includegraphics[width=\textwidth]{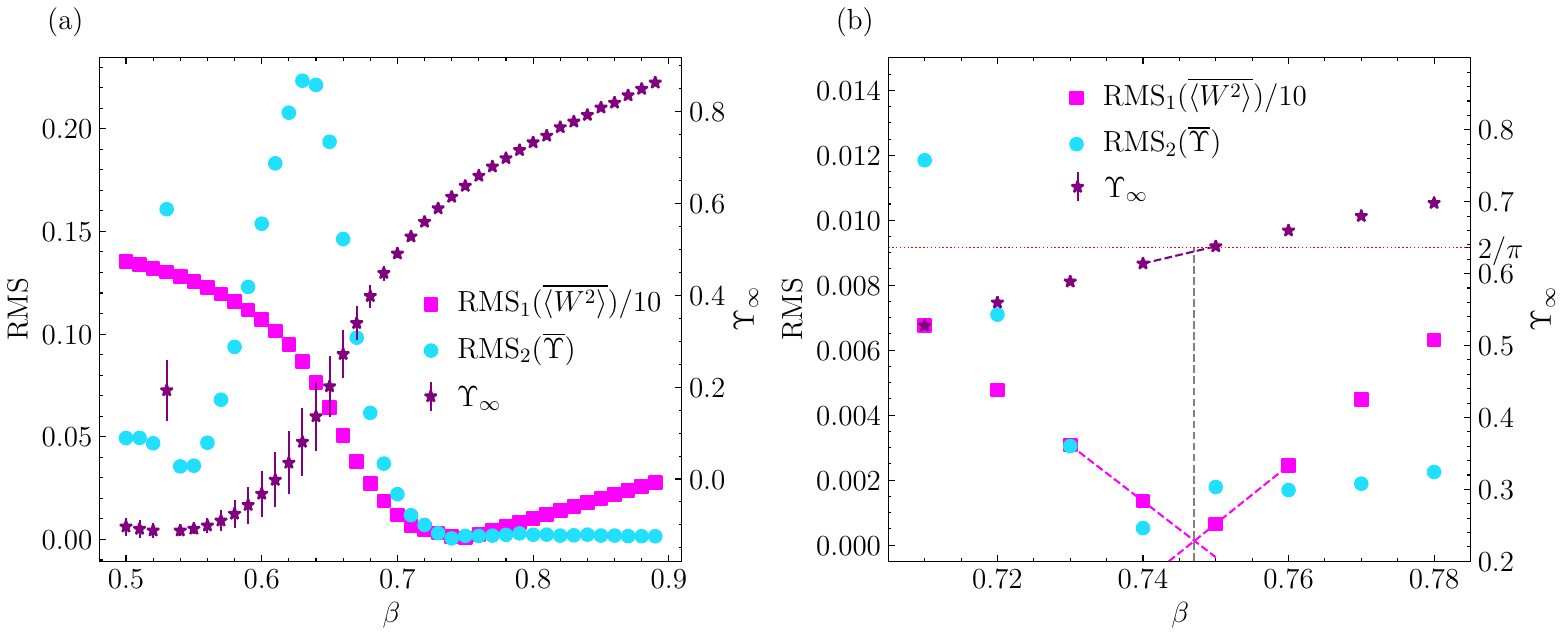}
        \caption{Same as Fig.~\ref{fig:RMSNL} but along the $\alpha = 2\beta / 17$ cut. (b) The critical point and stiffness jump are estimated using linear extrapolation.}
        \label{fig:RMSalmostClean}
\end{figure}

\begin{figure}
     \centering
         \includegraphics[width=0.85\textwidth]{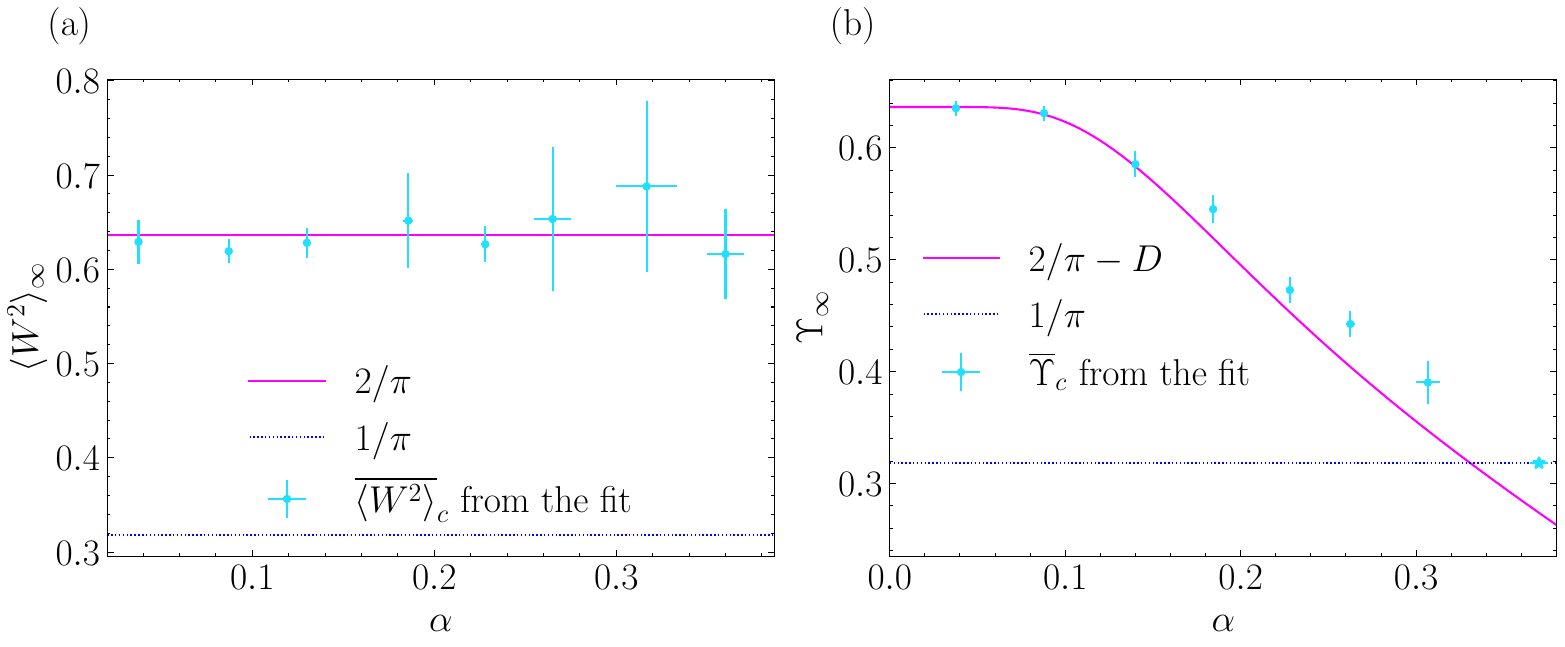}
        \caption{(a) Comparison of the numerically obtained jumps $\langle W^2 \rangle_\infty$, from the two-parameter fit, with $2/\pi$. The $1/\pi$ line is given for comparison with panel (b). (b) The non-universal jump of the helicity modulus at the critical line above the Nishimori line. The blue points are numerically obtained values of $\Upsilon_\infty$ from the scaling ansatz \eqref{eq:stiffness_FSSapp}. The magenta line is the prediction of the weak disorder field theory \eqref{eq:KcWeakDis} with the disorder strength $D$ given by the microscopic equation \eqref{eq:WDvar}.}
        \label{fig:jumps}
\end{figure}

To obtain a more stable procedure, we use the fact that the mean-square winding $\overline{\langle W^2\rangle}$ exhibits a universal jump of $2/\pi$ as predicted by the weak-disorder field theory result [see Eq.~\eqref{eq:uniWindJump}]. By fixing the jump, we simplify the problem to a single-parameter fit and obtain a more stable systematic procedure. A downside of using the mean-square winding is that the obtained numerical data at finite disorder is much noisier than that for the helicity modulus. However, since $\overline{\langle W^2 \rangle} = 2\overline{\Upsilon}$ everywhere on the Nishimori line [cf. Eq.~\eqref{eq:NLrelation}], we can fix the jump $\Upsilon_\infty = 1/\pi$ and use a single-parameter fit for $\Upsilon$. We observe that the positions of the minima of the RMS errors of the one- and two-parameter fits are consistent with each other on the Nishimori line (Fig.~\ref{fig:RMSNL}) and everywhere along the cuts $\beta = \lambda \alpha$ above the Nishimori line (see an example in Fig.~\ref{fig:RMSalmostClean}). By also implementing a two-parameter fit for $\overline{\langle W^2 \rangle}$, we compare the jump $\overline{\langle W^2\rangle}_\infty$ with $2/\pi$, as predicted by the weak-disorder field theory [Fig.~\ref{fig:jumps}(a)].

\begin{figure}
     \centering
         \includegraphics[width=0.9\textwidth]{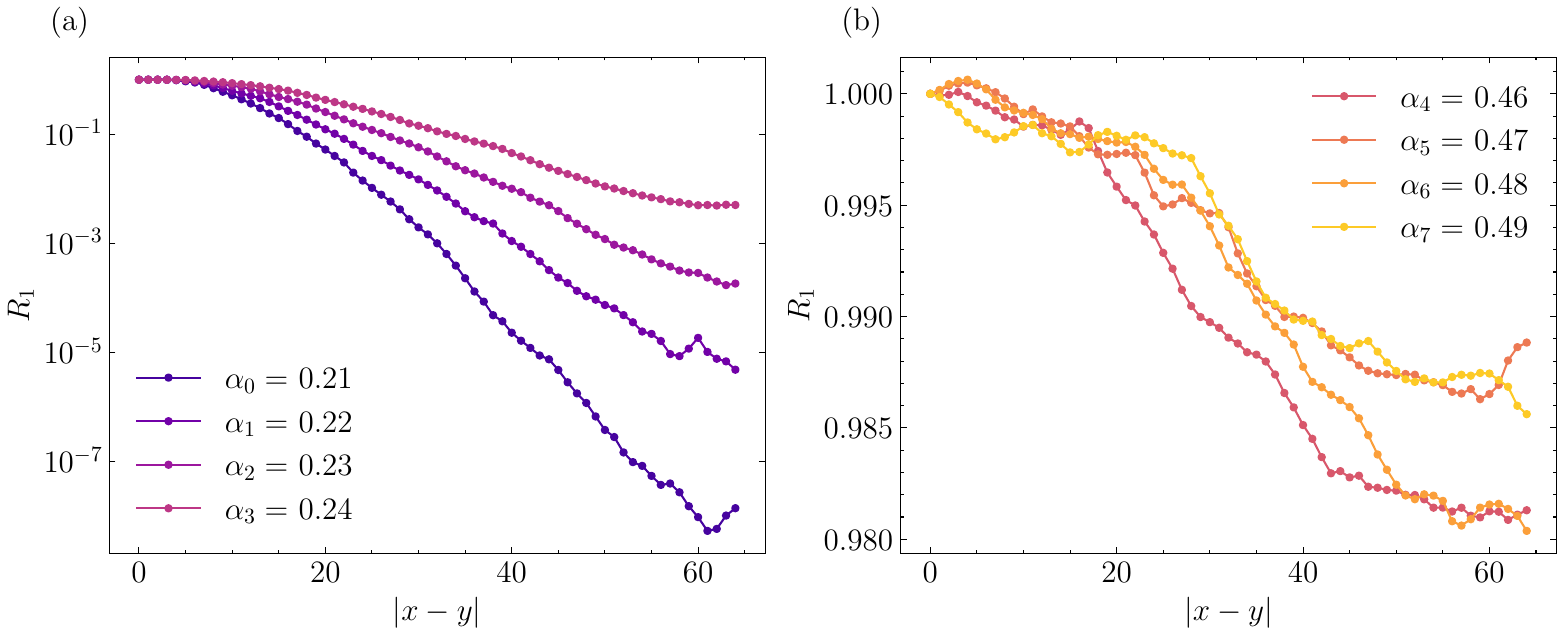}
        \caption{(a) The R\'enyi-1 correlator $R_1(x,y)$ in a system of linear size $L=128$, for decoherence strengths $\qty{\alpha_0, \alpha_1, \alpha_2, \alpha_3} = \qty{0.21, 0.22, 0.23, 0.24}$ below the critical point. The y-axis is in the log scale. (b) $R_{1}(x,y)$ for decoherence strengths $\qty{\alpha_4, \alpha_5, \alpha_6, \alpha_7} = \qty{0.46, 0.47, 0.48, 0.49}$ above the critical point.}
        \label{fig:R1plot}
\end{figure}

In the range of parameters where disorder is very weak, we slightly improve the precision by linearly extrapolating the RMS in the vicinity of the minima, e.g., see Fig.~\ref{fig:RMSalmostClean}. Our results for the critical temperature at very weak disorder are in agreement with \cite{jankeVillainMC1991}. 

Finally, Fig.~\ref{fig:jumps}(b) shows the dependence of the helicity modulus jump from the ansatz~\eqref{eq:stiffness_FSSapp} as a function of the disorder strength along the critical line. The numerical result is compared to the prediction of the weak disorder field theory, cf. Eq.~\eqref{eq:KcWeakDis}.

As mentioned in the main text, we also compute the R\'enyi-1 correlator signaling SWSSB, expressed as $\overline{\sqrt{\langle e^{i \theta_x} e^{-i \theta_y} \rangle}}$ on the Nishimori line [cf. Eq.~\eqref{eq:R1map}]. The numerical results below and above the error threshold are presented in Fig.~\ref{fig:R1plot}. As expected, we observe an exponential decay in the short-loop phase. The behavior in the long-loop phase is substantially longer ranged, but finite lattice sizes and disorder effects prevent us from determining the precise decay of the correlator. Our numerical results are therefore suggestive, albeit not conclusive, of (quasi-)SWSSB for $\alpha > \alpha_c$.

\subsection{Short-loop to loop-glass transition}

\begin{figure}
     \centering
         \includegraphics[width=\textwidth]{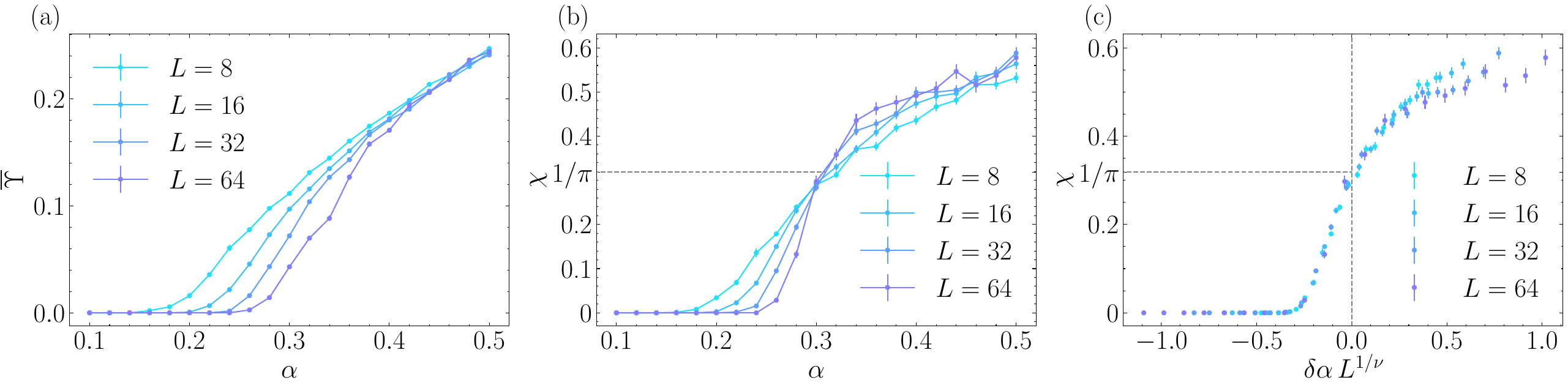}
        \caption{Numerical data on the $\beta = \alpha/2$ cut, for various system sizes. (a) Disorder-aveaged helicity modulus $\overline{\Upsilon}$. (b) Edwards-Anderson helicity modulus $\chi$. (C) Finite-size scaling collapse for $\chi$ with $\nu = 2.5$ and $\delta \alpha = \alpha - 0.307$.}
        \label{fig:AfNL2}
\end{figure}

\begin{figure}
     \centering
         \includegraphics[width=\textwidth]{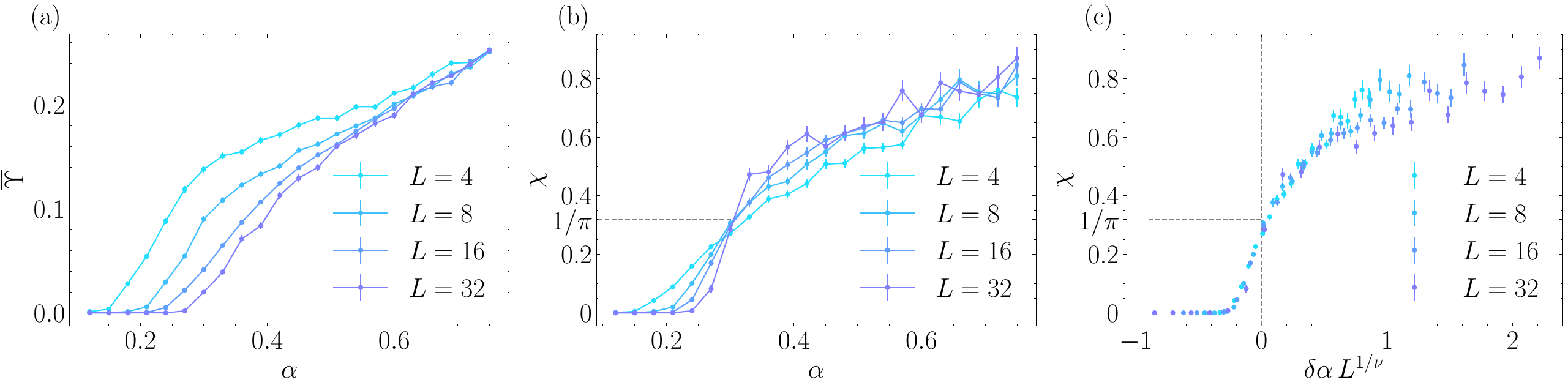}
        \caption{Numerical data on the $\beta = \alpha/3$ cut, for various system sizes. (a) Disorder-aveaged helicity modulus $\overline{\Upsilon}$. (b) Edwards-Anderson helicity modulus $\chi$. (C) Finite-size scaling collapse for $\chi$ with $\nu = 2.19$ and $\delta \alpha = \alpha - 0.295$.}
        \label{fig:AfNL6}
\end{figure}

Similar to the data along the $\beta = 0.15$ line, we observe suppression of the helicity modulus $\overline{\Upsilon}$ and growth of the Edwards-Anderson helicity modulus $\chi$ as a function of the system size on the cuts with $\lambda < 1$. We locate the boundary of the short-loop phase below the Nishimori line by studying the position of the crossing in $\chi$ and implementing a conventional scaling collapse with a finite critical exponent $\nu$. Interestingly, we find that it works decently at low temperatures with $\nu$ close to or the same as $\nu_{\rm MCF} \approx 2.19$ (see Figs.~\ref{fig:AfNL2} and \ref{fig:AfNL6} with data for the $\lambda = \qty{\frac{1}{2}, \frac{1}{3}}$ cuts; the data on the $\lambda = \frac{2}{3}$ is much less clear and is not presented here). Moreover, as in the zero-temperature transition, the crossing in $\chi$ appears to be close to $1/\pi$ along the critical line below the Nishimori point. 

A more detailed further investigation is necessary to verify its existence and characterize the loop-glass phase.

\section{Symmetry-Enriched Toric Code and Comparison of Decoding Thresholds}
As emphasized in the main text, measuring the physical charge of anyons in U(1) symmetry-enriched topological codes provides dramatically more information than measuring the topological charge of the anyons alone. Consequently, we expect the charge-informed optimal decoder to dramatically outperform the charge-agnostic optimal decoder. It is interesting to quantify the difference between these two decoders numerically, in order to get a sense for how much more information is provided by measuring the U(1) charges.

Towards this end, we consider here a $\mathbb{Z}_{2}$ toric code enriched with a U(1) symmetry, such that the $e$ anyons carry an integer charge. The generalization to $\mathbb{Z}_{p}$ toric codes is straightforward. Since the threshold for the $\mathbb{Z}_{2}$ toric code has been well-studied \cite{dennisTQM2002}, we can directly compare the thresholds of the charge-informed and charge-agnostic decoders.

Concretely, we consider a square lattice with quantum rotors on each site $i$ as in the main text. We additionally introduce a qubit with Pauli operators $X_{i \mu}, Z_{i \mu}$ on each link $(i \mu)$ of the lattice, and define the toric code stabilizers
\begin{equation}
    A_{i} \equiv \begin{tikzpicture}[baseline={(current bounding box.center)}]
        \draw[line width = 1.0pt, gray!50] (1, 0) -- (-1, 0);
        \draw[line width = 1.0pt, gray!50] (0, 1) -- (0, -1);
        \node at (0.5, 0) {\large $X$};
        \node at (-0.5, 0) {\large $X$};
        \node at (0, 0.5) {\large $X$};
        \node at (0, -0.5) {\large $X$};
        \node at (0, 0) {\large $i$}; 
    \end{tikzpicture}, \quad \quad B_{p} \equiv \begin{tikzpicture}[baseline={(current bounding box.center)}]
        \draw[line width = 1.0pt, gray!50] (0, 0) -- (0, 1.5) -- (1.5, 1.5) -- (1.5, 0) -- cycle;
        \node at (0, 0.75) {\large $Z$};
        \node at (0.75, 0) {\large $Z$};
        \node at (1.5, 0.75) {\large $Z$};
        \node at (0.75, 1.5) {\large $Z$};
        \node at (0.75, 0.75) {\large $p$};
    \end{tikzpicture} .
\end{equation}
Our system is initialized in the tensor product $\rho_{0} \equiv \dyad{0}^{\otimes N} \otimes \dyad{\text{TC}}$, where $\ket{\text{TC}}$ is a simultaneous eigenstate of each $A_{i}$ and each $B_{p}$. We then apply the following strongly symmetric channel on each link $i \mu$ of the lattice: 
\begin{equation}
    \mathcal{E}^{\text{TC}}_{i \mu}(\rho) = \sum_{k_{i\mu} \in \mathbb{Z}} p_{k_{i \mu}} e^{-i k_{i\mu} \Delta_{\mu} \hat{\varphi}_{i}} Z_{i \mu}^{k_{i \mu}} \ \rho \ Z_{i\mu}^{k_{i\mu}} e^{i k_{i \mu} \Delta_{\mu} \hat{\varphi}_{i}} ,
\end{equation}
where $p_{k_{i \mu}} \propto e^{-k_{i\mu}^{2}/2\alpha}$ as in the main text. The operator $Z_{i \mu}$ creates $e$ anyons at sites $i$ and $i + e_{\mu}$, while $e^{-i k_{i\mu} \Delta_{\mu} \hat{\varphi}_{i}}$ creates $k$ quasiparticles (quasiholes) at site $i$ ($i + e_{\mu}$). Since $Z_{i \mu}^{k_{i \mu}} = 1$ for even $k_{i \mu}$, each odd-integer U(1) charge is also an $e$ anyon, while even-integer charges are topologically trivial.

Whereas the charge-informed optimal decoder has access to the global charge distribution $n$, the charge-agnostic optimal decoder has access only to the locations of the $e$ anyons, i.e., it learns $n_{i} \mod 2$ on each site. In other words, the charge-agnostic optimal decoder is effectively attempting to decode a toric code under phase-flip decoherence with error probability
\begin{equation}
\label{eq:pz_alpha}
    p_{z}(\alpha) = \sum_{k_{i \mu} \in \text{Odd}} p_{k_{i \mu}} = \frac{\sum_{k_{i \mu} \in \text{Odd}} e^{-k_{i \mu}^{2} / 2\alpha}}{\sum_{k_{i \mu} \in \mathbb{Z}} e^{-k_{i\mu}^{2} / 2\alpha}} = \frac{\theta_{2}(0, e^{-2/\alpha})}{\theta_{3}(0, e^{-1/2\alpha})} ,
\end{equation}
where $\theta_{i}(z,\tau)$ are the Jacobi theta functions. 

As demonstrated in the main text, the charge-informed decoder exhibits an optimal decoding threshold at $\alpha_{c} \approx 0.37$. On the other hand, the charge-agnostic optimal decoder is that of the standard toric code under phase-flip errors, with an error threshold $p_{z}^{\text{Agnostic}} \approx 0.109$ corresponding to the critical disorder strength of the Nishimori random-bond Ising model \cite{dennisTQM2002,nishimoriBook2001}. We can compare this to the charge-informed decoding threshold by computing the effective phase-flip error probability at $\alpha = \alpha_{c}$:
\begin{equation}
    p_{z}(\alpha_{c}) \approx 0.34 ,
\end{equation}
which is almost three times larger than $p_{z}^{\text{Agnostic}}$. We therefore find that access to the physical U(1) charge of the anyons provides a dramatic enhancement to the optimal decoding threshold. 

It is also interesting to compare the minimal-cost flow threshold to the charge-agnostic optimal threshold. Using the value $\alpha_{\text{MCF}} \approx 0.265$ from the main text, we find
\begin{equation}
    p_{z}(\alpha_{\text{MCF}}) \approx 0.232 , 
\end{equation}
which is again a large enhancement over the charge-agnostic optimal threshold. In other words, even without knowledge of the underlying error probability distribution, access to the U(1) charges of errors allows for dramatically better error corection than any possible charge-agnostic decoder.

Alternatively, we can invert $p_z(\alpha)$ to find the error strength $\alpha = \alpha_c^{\text{Agnostic}}$ at which the charge-agnostic optimal decoder fails. Numerical inversion of Eq.~\eqref{eq:pz_alpha} gives
\begin{equation}
    \alpha_c^{\text{Agnostic}} \approx 0.179 ,
\end{equation}
which is substantially smaller than both the optimal threshold $\alpha_c$ and the MCF threshold $\alpha_c^{\text{MCF}}$.

\end{document}